\documentclass[preprint,12pt]{elsarticle}

\usepackage{graphicx}
\usepackage{xcolor}
\usepackage{natbib}
\usepackage{amsmath}
\usepackage{amssymb}
\usepackage{bm}
\usepackage{siunitx}

\makeatletter

\newcommand{\etal}{et al.\ }
\newcommand{\ie}{i.e.,\ }
\newcommand{\eg}{e.g.,\ }

\newcommand{\secref}[1]{\mbox{section \ref{#1}}}

\newcommand{\tabref}[1]{\mbox{table \ref{#1}}}

\newcommand{\equref}[1]{\mbox{equation (\ref{#1})}}

\newcommand{\figref}[2]{\mbox{figure \ref{#2}(#1)}}
\newcommand{\figrefSC}[1]{\mbox{Figure \ref{#1}}}
\newcommand{\figrefS}[1]{\mbox{figure \ref{#1}}}

\makeatletter
\def\@author#1{\g@addto@macro\elsauthors{\normalsize%
    \def\baselinestretch{1}%
    \upshape\authorsep#1\unskip\textsuperscript{%
      \ifx\@fnmark\@empty\else\unskip\sep\@fnmark\let\sep=,\fi
      \ifx\@corref\@empty\else\unskip\sep\@corref\let\sep=,\fi
      }%
    \def\authorsep{\unskip,\space}%
    \global\let\@fnmark\@empty
    \global\let\@corref\@empty  
    \global\let\sep\@empty}%
    \@eadauthor={#1}
}
\makeatother

\newcommand{\naokih}[1]{\textcolor{black}{#1}}

\journal{Computers \& Fluids}

\begin{document}
\begin{frontmatter}

\title{An Eulerian-based immersed boundary method for particle suspensions with implicit lubrication model}
\author{Naoki Hori$^1$}
\author{Marco E. Rosti$^{1,2}$\corref{cor1}}
\cortext[cor1]{Corresponding author: marco.rosti@oist.jp}
\author{Shu Takagi$^1$}
\address{$^1$ Department of Mechanical Engineering, The University of Tokyo, Tokyo, Japan}
\address{$^2$ Complex Fluids and Flows Unit, Okinawa Institute of Science and Technology Graduate University, 1919-1 Tancha, Onna-son, Okinawa 904-0495, Japan}

\begin{abstract}
We describe an immersed boundary method in which the fluid-structure coupling is achieved in an Eulerian framework. The method is an improved extension of the immersed boundary method originally developed by Kajishima \etal \citep{KAJISHIMA2001}, which accounts for the inertia of the fictitious fluid inside the particle volume and is thus able to reproduce the behaviour of particles both in the case of neutrally-buoyant objects and in the presence of density difference between the particles and the fluid. The method is capable to handle the presence of multiple suspended objects, \ie a suspension, by including a soft-sphere normal collision model, while the lubrication correction typically added to similar immersed boundary methods in order to capture the sub-grid unresolved lubrication force is here treated implicitly, \ie naturally obtained without any explicit expression, thus no additional computation is required. We show that our methodology can successfully reproduce the rheology of a particle suspension in a shear flow up to a dense regime (with a maximum particle volume fraction around $46\%$) without any additional correction force. The applicability of this methodology is also tested in a turbulent pressure-driven duct flow at high Reynolds number in the presence of non-negligible inertia and non-uniform shear-rate, showing good agreement with experimental measurements.
\end{abstract}


\end{frontmatter}

\section{Introduction}
Multiphase flows appear in many industrial processes, such as chemical engineering, food processing and resource mining. In such fields, a lot of work has been focused on the development of numerical methods which can successfully and reliably predict the behaviour of the dispersed phase and the modification induced on the carrier fluid. In general, the characteristics of the dispersed phases may vary widely. If the length-scale of the objects is very small, we can often model their behaviour as a continuum by modifying the equations describing the dynamics of the fluid, \eg the effect of polymer suspensions in a Newtonian carrier fluid. In other cases, the dispersed phase has length-scales that are comparable to the macroscopic ones governing the problem, and their full dynamics must be properly captured by the numerical method. In such cases, all the complex features of the dispersed phase need to be considered, \eg shape, deformability, coalescence, breakup. In the present work we focus on dispersed objects which are rigid, and we propose a numerical method to simulate rigid particle-laden flows as found in many applications.

In the past, several methods were developed to simulate the flows laden with rigid particles.  One of the most successful approach is the so-called immersed boundary method (IBM), in which the suspended object is described by a volume force applied to the fluid phase which restores the proper boundary conditions at the solid surface. This method was originally designed by Peskin \citep{PESKIN2002} to study the biological flow inside the human heart, and is now extensively used to simulate the flows in complex geometry, see \eg Fadlun \etal \citep{FADLUN2000} and Das \etal \citep{DAS2018}. The immersed boundary method has been also applied to the particle-laden flows where the motion of the suspended particles is coupled to the fluid governing equations. Uhlmann \citep{UHLMANN2005} developed an IBM to simulate the particulate flows, a so-called direct-forcing IBM, in which the no-slip boundary conditions on the particle surface is imposed directly. The velocity difference between the particle and the fluid velocity is used to evaluate the force and torque acting on the particles and to restore the proper boundary condition on the fluid. By positioning some calculation points (Lagrangian points) on the surface of the particles in addition to the normal Eulerian mesh and communicating the velocity and force information between them, the method describes the particles surface effects explicitly. Although still categorized as direct-forcing IBM, Kajishima \etal \citep{KAJISHIMA2001} proposes a different approach to describe the interaction between the fluid and the particles: the authors describe the particle in terms of the volume fraction occupied on the Eulerian grid of the fluid. This procedure prescribe the no-slip boundary condition not on the Lagrangian surface but in the Eulerian cells containing it. Because of this, Lagrangian points are not used, which is a preferable feature from the computational cost perspective; indeed, this massively simplify the numerical scheme and its parallelisation procedure allowing for fast computation and easy migration towards the rapidly growing graphics processing unit (GPU) computations.  The method proposed by Kajishima \etal \citep{KAJISHIMA2001} is often classified as immersed body method \cite{maxey_2017v}, a class of methods which includes among others the so-called smoothed profile method  originally developed by Nakayama and Yamamoto \cite{nakayama_yamamoto_2005k, yamamoto_molina_nakayama_2021m},  and later extended by Luo \etal \cite{ luo_maxey_karniadakis_2009i}. Apart from the IBM, several other techniques have been proposed and used in the past; among those using a fictitious domain approach,  techniques worth mentioning are the distributed Lagrange multiplier method first developed by Glowinski \etal \cite{glowinski_pan_hesla_joseph_1999c}, where the fluid equations are solved in the whole domain and coupled with the particle ones in a monolithic form,  the Physalis method developed by Prosperetti and Oguz \cite{prosperetti_oguz_2001i} where the flow near the surface of a particle is  represented by the solution of the Stokes flow,  assuming that such flow is dominated by viscous forces even at finite Reynolds number. A full classification of the methods and a description of their main characteristics and differences can be found in the review by Maxey \cite{maxey_2017v}.

What discussed above mainly refers to single particles immersed in a fluid, but the interactions between particles are inevitable when considering full suspensions. Three main interactions exist among particles immersed in a fluid (or between a particle and a wall): the hydrodynamic long-range interaction, and the short-range collision and lubrication. While the long-range one is naturally captured by all the immersed boundary methods, the short-range ones are not, with particles penetrating into each other because of the implicit treatment of their surface and because of the impossibility of fully resolving the fluid lubrication force for gaps between particles smaller than the grid-size. These issues lead to the unrealistic description of the phenomena, and sometimes the divergence of the simulations. Inter-particle penetration can be easily avoided by using a proper collision model; because of the stiffness of the collision problem (high forces in short times), the so-called soft-sphere collision model was first proposed by Cundall and Strack \citep{CUNDALL1979} and is regarded as a useful model to couple with the IBM. In this model, small inter-particle penetration are allowed and the amount of penetration is used to evaluate the amplitude of the collision force. The lubrication force is another short-range hydrodynamic effect, which arises when two objects get close. Brenner \citep{BRENNER1961} analytically shows that when a spherical particle approaches or departs from a wall it is subject to a force which opposes to its motion. Common numerical schemes cannot properly capture this short-range hydrodynamic force because of the finite size of the grid and the consequent lack of resolution as the distance between the objects reduces. To overcome this problem, sub-grid forces are usually added and several models have been proposed \cite{KULKARNI2008, BREUGEM2010, BREUGEM2012}. Although these models are able to properly reproduce the correct dynamics in several applications \citep{SOLDATI2009, PICANO2013, PICANO2015, ZADE2019}, they are strongly dependent on the numerical method used to describe the particle, they rely on the tuning of several model parameters, they introduce non-negligible additional computational cost, and sometimes include some ambiguity. In the present work, we still rely on the soft-sphere collision model to avoid inter-particle penetration but we propose a different approach to capture the lubrication force: our method is implicit in the sense that no additional force is added to capture the proper suspension dynamics and is based on the exploit of the unresolved lubrication force naturally captured by the immersed boundary method.

This manuscript is organized as follows: in section 2 we describe the mathematical formulation and numerical methodology used to simulate a single object immersed in a fluid flow using an Eulerian-based IBM. We also show the validity of the method by comparing our results with several experimental and numerical results available in the literature. In section 3, we describe how to handle the interaction of multiple particles, especially how we implicitly evaluate the correct lubrication force. We test the method by showing the rheological property of a suspension in a laminar shear flow. Next, the results obtained with the whole methodology in a moderately high Reynolds number turbulent pressure driven flow are shown. Finally, a summary of the main conclusions is reported in section 4.

\section{The Eulerian-based immersed boundary method}
\subsection{Mathematical formulation}
We consider an incompressible fluid with immersed particles. The fluid is governed by the momentum conservation,
\begin{equation} \label{eq:momentum_conservation}
  \rho^f \left( \frac{\partial u_i}{\partial t} + \frac{\partial u_i u_j}{\partial x_j} \right) = \frac{\partial \sigma_{ij}}{\partial x_j} + \rho^f g_i + \rho^f a_i,
\end{equation}
and the incompressibility constraint,
\begin{equation} \label{eq:incompressibility_constraint}
  \frac{\partial u_i}{\partial x_i} = 0,
\end{equation}
where the Einstein notation is used in the subscripts. $\rho^f$ is the density of the fluid, $u_i$ the velocity, $\sigma_{ij}$ the Cauchy stress tensor, $g_i$ the gravitational acceleration, and $\rho^f a_i$ an external body force which is imposed to couple the particle interaction described later. If we assume that the fluid is Newtonian, the Cauchy stress tensor can be defined as
\begin{equation}
  \sigma_{ij} = -p \delta_{ij} + 2 \mu^f \mathcal{D}_{ij},
\end{equation}
where $p$ is the pressure, $\delta_{ij}$ the Dirac delta, $\mu^f$ the fluid viscosity, and $\mathcal{D}_{ij} = \left( \partial u_i / \partial x_j + \partial u_j / \partial x_i \right)/2$ the strain rate tensor.

When we assume that a particle is rigid, the velocity at an arbitrary point $X_i$ inside the particle, $U_i \left( X_i \right)$, can be described using the translational velocity $U_i^{c}$ and the rotational velocity $\Omega_i^{c}$ of its center as
\begin{equation}
  U_i = U_i^{c} + \epsilon_{ijk} \Omega_j^{c} r_k,
\end{equation}
where $\epsilon_{ijk}$ is the Levi-Civita permutation symbol and $r_k$ the vector going from the center of the particle to $X_i$. The time evolution of the center translational and rotational velocities are governed by the Newton-Euler equations,
\begin{subequations} \label{eq:newtoneuler}
\begin{align}
  m^p \frac{dU_i^{c}}{dt} &= \oint_{\partial \mathcal{V}^p} \sigma_{ij} n_j dA + F_i^{\text{ext}}, \\
  \mathcal{I}^p \frac{d\Omega_i^{c}}{dt} &= \oint_{\partial \mathcal{V}^p} \epsilon_{ijk} r_j \sigma_{kl} n_l dA + T_i^{\text{ext}},
\end{align}
\end{subequations}
where $m^p$ and $\mathcal{I}^p$ are the mass and the moment of inertia of the particle, respectively. For a spherical particle whose radius is $r^p$, $m^p$ and $\mathcal{I}^p$ are equal to $4/3 \pi \rho^p {r^p}^3$ and $2/5 m^p {r^p}^2$, where $\rho^p$ is the density of the particle. $\sigma_{ij}$ is the same Cauchy stress tensor described in \equref{eq:momentum_conservation}, $n_i$ the vector normal to the particle at the point and $F_i^{\text{ext}}$ and $T_i^{\text{ext}}$ are the total external force and torque, \eg gravity. On the surface of a particle, the no-slip and no-penetration conditions are applied, \ie the fluid velocity on the surface equals to the local particle velocity $U_i=u_i$.

\subsection{The immersed boundary method} \label{subsec:numerical_procedures}
In the original immersed boundary method developed by Kajishima \etal \citep{KAJISHIMA2001}, the presence of the particle is accounted for by means of a body force $\rho^f a_i$ added to the momentum equation where the acceleration $a_i$ is defined as
\begin{equation} \label{eq:particle_bc}
  a_i = \alpha \frac{U_i - u_i}{\Delta t},
\end{equation}
where $\alpha$ is the particle volume fraction in the considered cell, $U_i$ and $u_i$ the particle and fluid velocities, and $\Delta t$ the time step. Note that, the body force $\rho^f a_i$ is zero when $\alpha=0$ (when the grid cell is totally filled by the fluid), and not null otherwise (when the grid cell is partially or totally occupied by the particle). In the particle equations, instead of evaluating the Cauchy stress tensor term of \equref{eq:newtoneuler} directly on the particle surface, Kajishima \etal suggested to use the above volume acceleration term, \ie
\begin{subequations}
\begin{align}
  m^p \frac{dU_i^{c}}{dt} &= -\rho^f \int_{\mathcal{V}^p} a_i d\mathcal{V} + F_i^{\text{ext}}, \\
  \mathcal{I}^p \frac{d\Omega_i^{c}}{dt} &= -\rho^f \int_{\mathcal{V}^p} \epsilon_{ijk} r_j a_k d\mathcal{V} + T_i^{\text{ext}}.
\end{align}
\end{subequations}
Uhlmann \citep{UHLMANN2005} found that the net body force $-\rho^f \int_{\mathcal{V}^p} a_i d\mathcal{V}$ exhibits a spurious oscillatory behaviour when simulating a forced-oscillating cylinder. Although this phenomena still exists in the current IBM (see \ref{app:forced} for further details), we have never observed this behaviour when the object is freely moving or stationary. Also, this oscillations can be suppressed by adopting more smoothed interface digitiser. For simplicity, here we limit our discussion to the original digitiser \citep{YUKI2007}.

Kajishima \etal \citep{KAJISHIMA2001} added the body force $\rho^f a_i$ in the momentum equation at the end of the time advancement scheme, thus effectively enforcing the particle rigid body motion inside the particle. Although formally correct, such procedure has the drawback that the incompressibility constraint (\equref{eq:incompressibility_constraint}) is in general not exactly satisfied because the divergence free velocity of the fluid is distorted by the particle rigid body motion. Also, the imposition of the rigid body motion and its use in the evaluation of the force and torque integrals controlling the particle dynamics lead to a singularity in the equations when the density ratio of the two phases $\rho^p/\rho^f$ is close to $1$. Uhlmann \citep{UHLMANN2005} reported that the lower limit of $\rho^p/\rho^f$ for which a stable solution can be obtained is around $2$, while Kajishima and Takiguchi \citep{KAJISHIMA2002} and Iwata \etal \citep{IWATA2010} found that the limit is around $1.6$ when using the second-order Adams-Bashforth scheme and $1.1$ with the third-order Runge-Kutta scheme. In order to resolve these two issues, we adapt the concepts described by Breugem \citep{BREUGEM2012} for a Lagrangian immersed boundary method to our Eulerian formulation. First, the divergence free of the velocity field can be recovered by changing the order in which the fluid and particle equations are solved and coupled together; in particular, we enforce the divergence free of the velocity after adding the immersed boundary force in the momentum equation, as will be discussed in the next section in more details. Because of this change, the fluid velocity inside the particle is not overridden with the rigid body motion and we can now take into account its motion and inertia. This correction is equivalent to applying the body force $\rho^f a_i$ only in the cells where the interface between the solid and fluid is present, \ie where $0<\alpha<1$. By doing so, the fluid inertia of the fluid inside the particle can be accounted for in the Newton-Euler equations as
\begin{subequations} \label{eq:newtoneuler2}
\begin{align}
  m^p \frac{dU_i^{c}}{dt} &= -\rho^f \oint_{\partial \mathcal{V}^p} a_i dA + \rho^f \frac{d}{dt} \int_{\mathcal{V}^p} u_i d\mathcal{V} + F_i^{\text{ext}}, \\
  \mathcal{I}^p \frac{d\Omega_i^{c}}{dt}   &= -\rho^f \oint_{\partial \mathcal{V}^p} \epsilon_{ijk} r_j a_k dA + \rho^f \frac{d}{dt} \int_{\mathcal{V}^p} \epsilon_{ijk} r_j u_k d\mathcal{V} + T_i^{\text{ext}},
\end{align}
\end{subequations}
and \equref{eq:particle_bc} is modified as
\begin{equation}
  a_i = \mathcal{H} \left( 1 - \alpha \right) \alpha \frac{U_i -u_i}{\Delta t},
\end{equation}
where $\mathcal{H} \left( x \right)$ is the step function which becomes $1$ when $x$ is positive and $0$ otherwise. The interested reader is referred to the work by Breugem \citep{BREUGEM2012} for the full analytical derivation of these equations.

The above procedure requires the determination of the volume fraction occupied by the solid object in each cell $\alpha$, which is generally a time-consuming task since the procedure is called several times. Tsuji \etal \citep{TSUJI2003} proposed the so-called subdivision volume counting method, while Kempe \etal \citep{KEMPE2009} computed it using a level-set function. Although second-order accuracy can be achieved with these methods, they are time consuming and we opted to use the method by Yuki \etal \citep{YUKI2007} which has an extremely low computational cost. In particular, the authors proposed to calculate the volume fraction of a particle by assuming a sigmoid-like surface as
\begin{subequations} \label{eq:surface}
  \begin{align}
    \alpha &= \frac{1}{2} \left[ 1-\tanh \left( \frac{\delta s}{\sigma \lambda \Delta} \right) \right], \\
    \lambda &= | n_x | + | n_y | + | n_z |, \\
    \sigma &= 0.05 \left( 1-\lambda^2 \right) + 0.3,
  \end{align}
\end{subequations}
where $\delta s$ is a signed distance from the cell centre to the surface element, $n_i$ the normal vector, and $\Delta$ is the reference mesh size. By using this digitaliser, the volume of the object can be recovered with errors below $0.5\%$ for a resolutions equal to $D_p/\Delta=16$. Note that, in the two-dimensional case we set $\left| n_z \right|=0$ in the expression above.

Note that, Bigot \etal \citep{BIGOT2014} proposed a similar strategy to simulate immersed objects in constant/stratified density fields. In their formulation, however, the aforementioned singularity issue was not solved and their method was thus not able to consider neutrally-buoyant objects.

The main advantage of this IBM is the absence of any Lagrangian points, see \eg Peskin \citep{PESKIN2002}. While the use of Lagrangian points has the merit of easily and explicitly describing the interface, it requires additional computational cost (both in terms of time and memory) to exchange the information with Eulerian grids in the so-called interpolation and spreading procedures, and in general its parallelisation is complicated (see \eg Uhlmann \citep{UHLMANN2004}). On the other hand, in an Eulerian framework the above mentioned additional computational cost is absent and the parallelisation of the algorithm is straightforward. Also, the smooth treatment of the surface due to the use of the sigmoid function contributes to the stability of the coupling with the fluid and is also consistent with other Eulerian techniques, such as the volume of fluid method \citep{II2012, ROSTI2019}, and thus a preferable option when studying three-phase flows.

\subsection{Numerical discretisation} \label{subsec:time_integration}
Next, we describe how the above equations are solved numerically. The second-order central finite difference scheme is used for the spatial dicretisation and the second-order Adams-Bashforth scheme  coupled with a fractional step method for the time advancement. In particular, the fluid solver is based on the classical Simplified Marker and Cell (SMAC) method \citep{HARLOW1965}, which splits \equref{eq:momentum_conservation} into two steps,
\begin{subequations}
  \begin{align}
    u_i^* &= u_i^n + \Delta t \left( -\frac{1}{\rho^f}\frac{\partial p^n}{\partial x_i} + \frac{3}{2} \left( rhs \right)_i^n - \frac{1}{2} \left( rhs \right)_i^{n-1} \right), \label{eq:smac1}\\
    u_i^{n+1} &= u_i^{**}-\frac{\Delta t}{\rho^f} \left( \frac{\partial p^{n+1}}{\partial x_i} - \frac{\partial p^n}{\partial x_i} \right), \label{eq:smac2}
  \end{align}
\end{subequations}
where $\left( rhs \right)_i$ is the sum of the advection, diffusion and gravity terms, \ie $\left( rhs \right)_i = -\frac{\partial u_i u_j}{\partial x_j} + \frac{1}{\rho^f}\frac{\partial 2 \mu^f \mathcal{D}_{ij}}{\partial x_j} + g_i$. $u_i^*$ is the first intermediate velocity, which is non-divergence free and is used to exchange the momentum with a particle, while $u_i^{**}$ is the second intermediate velocity obtained after the imposition of the immersed boundary force
\begin{equation} \label{eq:particle_bc_2}
  \rho^f a_i^{n+1} = \rho^f \mathcal{H} \left( 1 - \alpha^n \right) \alpha^n \frac{\left( U_i^n - u_i^* \right)}{\Delta t}
\end{equation}
as
\begin{equation}
  u_i^{**} = u_i^* + \frac{\Delta t}{2} \left( a_i^{n+1} + a_i^n \right),
\end{equation}
where the second-order Crank-Nicolson scheme is used here. Finally, the pressure $p^{n+1}$ in \equref{eq:smac1} is found by solving the pressure correction equation
\begin{equation} \label{eq:poisson}
  \frac{\partial^2p^{n+1}}{\partial x_i \partial x_i} = \frac{\rho^f}{\Delta t} \frac{\partial u_i^{**}}{\partial x_i} + \frac{\partial^2p^n}{\partial x_i \partial x_i}.
\end{equation}
The acceleration in \equref{eq:particle_bc_2} is used to update the particle center translational and angular velocities by solving \equref{eq:newtoneuler2} as
\begin{subequations} \label{eq:newtoneuler3}
  \begin{align}
    U_i^{p_c,n+1}      &= U_i^{p_c,n}      + \frac{1}{m^p}           \left[ \frac{\Delta t}{2} \left( A_i^{n+1} + A_i^n \right) + \Delta I_i + F_i^{\text{ext}} \Delta t \right], \\
    \Omega_i^{p_c,n+1} &= \Omega_i^{p_c,n} + \frac{1}{\mathcal{I}^p} \left[ \frac{\Delta t}{2} \left( B_i^{n+1} + B_i^n \right) + \Delta J_i + T_i^{\text{ext}} \Delta t \right],
  \end{align}
\end{subequations}
where we have defined
\begin{subequations}
  \begin{align}
    A_i^{n} &= -\rho^f \sum_{\partial \mathcal{V}^p} a_i^{n} \Delta \mathcal{V}, \\
    B_i^{n} &= -\rho^f \sum_{\partial \mathcal{V}^p} \epsilon_{ijk} r_j a_k^{n} \Delta \mathcal{V}, \label{eq:torque_tc}
  \end{align}
\end{subequations}
and
\begin{subequations}
  \begin{align}
    \Delta I_i &= \rho^f \left( \sum_{\mathcal{V}^p} u_i^{*} - \sum_{\mathcal{V}^p} u_i^n \right) \Delta \mathcal{V}^p, \\
    \Delta J_i &= \rho^f \left( \sum_{\mathcal{V}^p} \epsilon_{ijk} r_j u_k^{*} - \sum_{\mathcal{V}^p} \epsilon_{ijk} r_j u_k^n \right) \Delta \mathcal{V}^p.
  \end{align}
\end{subequations}
Here, $\Delta \mathcal{V}$ is the volume of a cell, equal to $\Delta x \Delta y$ in two dimensions and $\Delta x \Delta y \Delta z$ in three dimensions. Using the updated particle center and angular velocities we can update the center position as
\begin{equation}
  X_i^{p_c,n+1} = X_i^{p_c,n} + \frac{\Delta t}{2} \left( U_i^{p_c,n+1} + U_i^{p_c,n} \right).
\end{equation}
Note that, the use of a CN scheme in the update of particle position and velocity has been first proposed by Takeuchi \etal \cite{TAKEUCHI2015} in the framework of the present IBM.
Also note that, in a similar way we can also obtain the rotational angle, but this is not necessary for circles and spheres other than for visualisation purposes. Although the update procedure is completed, in order to further increase the stability of the scheme the particle velocity term $U_i^n$ in \equref{eq:particle_bc_2} can be treated implicitly by using the particle position at the ${n+1}$ step. To achieve this, we can include an iteration in the above procedures repeated until the final particle position converges, \ie
\begin{subequations} \label{eq:ibm_integrate}
  \begin{align}
    &u_i^* = u_i^n + \Delta t \left( - \frac{1}{\rho^f} \frac{\partial p^n}{\partial x_i} + \frac{3}{2} \left( rhs \right)_i^n - \frac{1}{2} \left( rhs \right)_i^{n-1} \right), \label{eq:prediction} \\
    &\text{do} \nonumber \\
    &\ \ \ \ a_i^{k+1} = \mathcal{H} \left( 1 - \alpha^k \right) \alpha^k \frac{\left( U_i^k - u_i^* \right)}{\Delta t}, \\
    &\ \ \ \ U_i^{p_c,k+1} = U_i^{p_c,n} + \frac{1}{m^p} \left[ \frac{\Delta t}{2} \left( A_i^{k+1}+A_i^{n} \right) + \Delta I_i + F_i^{\text{ext}} \Delta t \right], \\
    &\ \ \ \ \Omega_i^{p_c,k+1} = \Omega_i^{p_c,n} + \frac{1}{\mathcal{I}^p} \left[ \frac{\Delta t}{2} \left( B_i^{k+1}+B_i^{n} \right) + \Delta J_i + T_i^{\text{ext}} \Delta t \right], \\
    &\ \ \ \ X_i^{p_c,k+1} = X_i^{p_c,n} + \frac{\Delta t}{2} \left( U_i^{p_c,k+1} + U_i^{p_c,n} \right), \\
    &\ \ \ \ k = k+1, \nonumber \\
    &\text{while} \| X_i^{p_c,k+1} - X_i^{p_c,k} \| < \epsilon, \nonumber
  \end{align}
\end{subequations}
where $k$ is the sub-iteration counter, and $\epsilon$ is a sufficiently small number. Note that, at the beginning of the sub-iteration loop ($k=1$) the variables at $k$ are regarded as those with $n$, while at convergence the variables at ${k+1}$ are regarded as the final ${n+1}$ step variables. Thanks to the sub-iteration procedure, we can now use the second-order Crank-Nicolson scheme to evaluate the acceleration terms ($A_i$ and $B_i$), the velocities and the positions. It should be noted that the above method recovers the one proposed by Takeuchi \etal \citep{TAKEUCHI2015} when only one iteration is used, \ie a predictor-corrector scheme, and that Breugem \citep{BREUGEM2012} suggested a similar algorithm. Breugem also found that the optimal iteration number $k_{\text{max}}$ is equal to $2$ as a balance between stability and cost. In our method, the cost of the sub-iteration is negligible compared to the fluid solution because of the Eulerian treatment of the particles which requires little communication among different processors in the sub-iteration, and thus we can iterate until the fully converged state is achieved.

The implementation of the above algorithm is based on the use of the Message Passing Interface (MPI) library for parallelisation and on the Fastest Fourier Transform in the West (FFTW) library to solve the pressure Poisson equation.

\begin{figure}
\centering
\includegraphics[width=\textwidth]{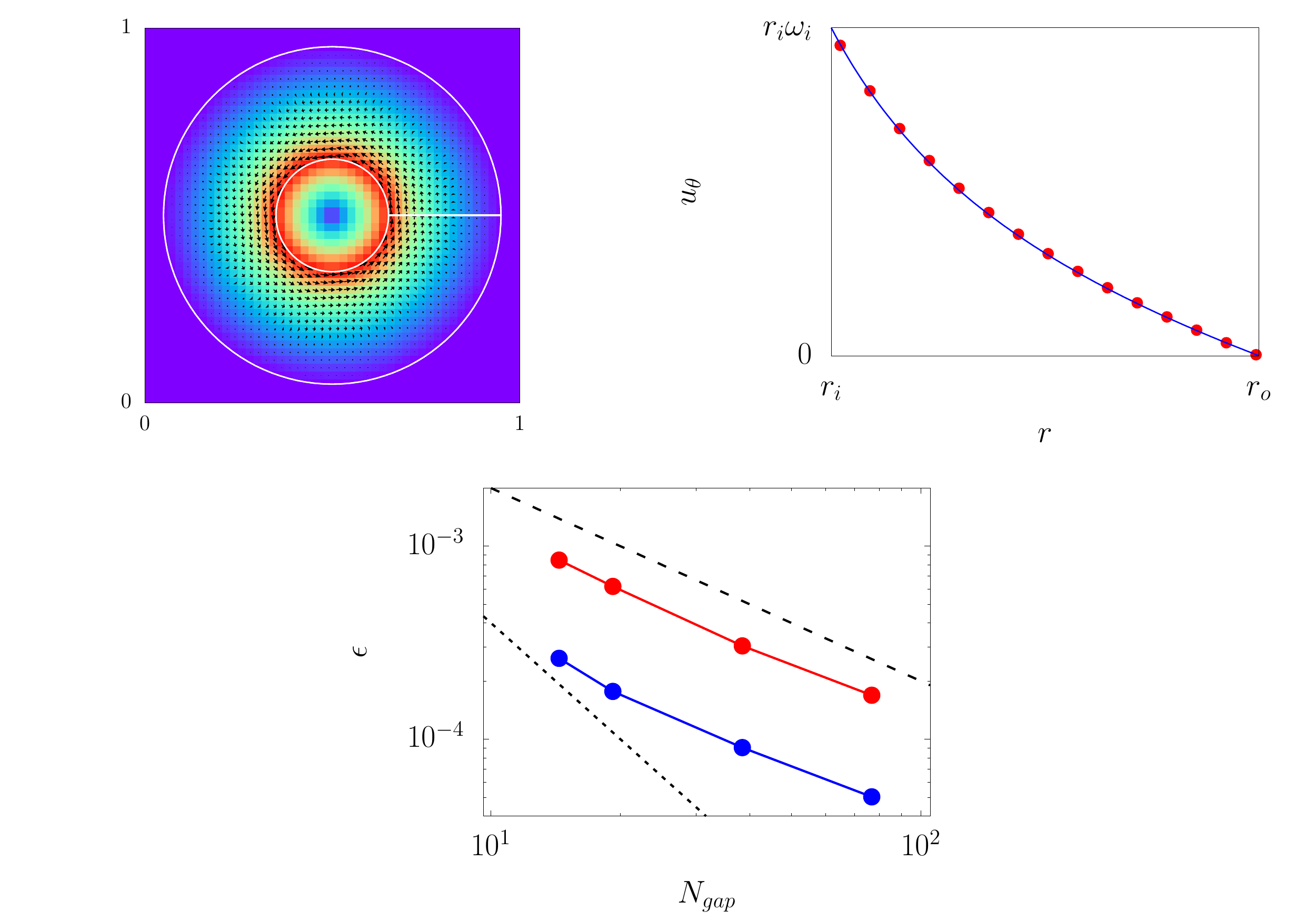}
  \caption{
    ($a$) A visualization of the flow field with $N_{gap} \approx 15$,  with the color contour denoting the magnitude of the velocity field, and the arrows showing the velocity directions.  The two white circles indicate the inner and outer cylinders, and the velocity on the white horizontal line is plotted in ($b$) as red dots, whereas the blue line indicates the analytical solution (equation \eqref{eq:tc_analytical}).
    \naokih{($c$)} Convergence errors as a function of the number of grid points in the gap $N_{gap}$: \naokih{(red)} azimuthal velocity, and \naokih{(blue)} torque.
    \naokih{Black dashed and dotted lines denote the first and second-order accuracy in space, respectively.}
  }
  \label{fig:TC}
\end{figure}

\subsection{Results and validations for a single immersed object}
In this section we verify the validity of our numerical scheme and immersed boundary method by studying four different problems. First, we study the flow between two concentric rotating disks; next, we consider a two-dimensional shear flow and a two-dimensional pressure driven flow and study the migration of a circular and neutrally buoyant rigid cylinder, and finally, we consider the gravity-driven sedimentation of a rigid sphere with a non-unitary density ratio in a three-dimensional box.

\subsubsection*{Taylor-Couette flow}
In order to evaluate the spatial convergence of the current algorithm, we start by analysing the flow between two concentric disks whose radii are $r_i$ and $r_o$, respectively. We consider a square domain of unit length, in which two coaxial disks with $r_i = 0.15$ and $r_o = 0.45$ are located. The inner disk rotates at a fixed angular velocity $\omega_i = 1$, whereas the outer disk is fixed in space, giving a Reynolds number $Re \equiv \rho r_i \omega_i \left( r_o - r_i \right) / \mu = 0.9$.
\naokih{
  We enforce on the four domain boundaries the no-slip and no-penetration conditions for the velocity,  and the Neumann condition for the pressure.
  Note that, we have also tried periodic boundary conditions and confirmed the negligible effect on the following discussion.
}
The flow reaches a steady state after a sufficiently long time, with the velocity field shown in figure \ref{fig:TC}($a,b$). In order to quantify the error, we compare the azimuthal velocity profile $u_{\theta} \left( r \right)$ with the analytical solution given by
\begin{equation} \label{eq:tc_analytical}
  u_{\theta} \left( r \right) = -\frac{\eta^2 \omega_i}{1-\eta^2} r + \frac{r_i^2 \omega_i}{1-\eta^2} \frac{1}{r},
\end{equation}
where we defined the curvature $\eta$ as $\equiv r_i / r_o$.
\naokih{
  In addition, we quantify the error with respect to the torque $T$ needed to keep the inner disk rotating, whose analytical solution is
  \begin{equation}
    T = \frac{4 \pi \mu r_i^2 \omega_i}{1 - \eta^2}.
  \end{equation}
  We note that the error of the velocity field is evaluated by computing the $L^2$ norm inside the flow region, i.e., in the computational nodes where the radial position $r$ satisfies $r_i < r < r_o$.
  Velocities defined at the edges of the Cartesian staggered mesh are first interpolated to the corresponding cell center, which are then converted to the used cylindrical coordinate to be compared with the above analytical solution.
  Regarding the torque, the the $L^1$ norm of \eqref{eq:torque_tc} is considered to investigate the error.
  In figure \ref{fig:TC}($c$), we show the errors of the two quantities as a function of the number of grid points in the gap $N_{gap}$, where we observe that both quantities exhibit approximately a \naokih{first}-order accuracy in space.
}

\begin{figure}[t]
  \centering
\includegraphics[width=0.49\textwidth]{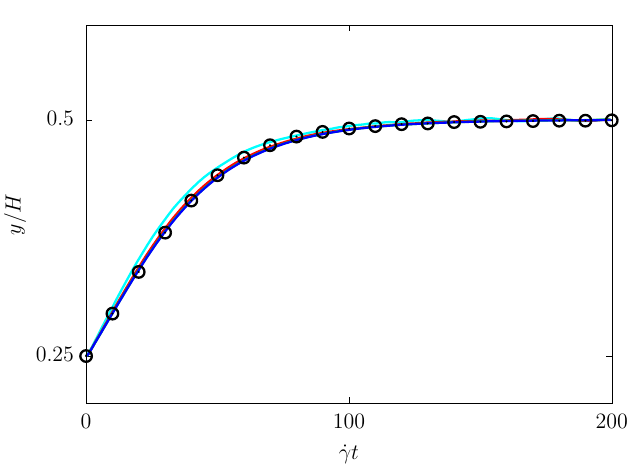}
  \caption{
    The particle migration in a shear flow and the effect of the grid resolution $D_p/\Delta$. The $x$ axis represents the time normalised by $H/U_{\text{wall}}=1/\dot{\gamma}$, while the $y$ axis the wall-normal position of the particle center normalised by the channel width $H$. Different colors are used to distinguish the different grid resolutions $D_p/\Delta$: (cyan) $8$, (red) $16$, (green) $32$, and (blue) $48$, while the black points are the results by Feng \etal \citep{FENG1994}. 
  }
  \label{fig:feng}
\end{figure}

\subsubsection*{Lateral migration of a circular cylinder in a shear flow} \label{subsubsec:cylinder_in_shear}
We now verify the validity of our IBM for suspended particles by showing the results of the lateral migration of a cylinder in a two-dimensional shear flow. Cox and Brenner \citep{COX1968} and Ho and Leal \citep{HO1974} performed a perturbation analysis and found that the stable position of the cylinder is the middle of the channel. However, their analysis is only valid when the Reynolds number is sufficiently smaller than $D_p/H$, where $D_p$ and $H$ are the particle diameter and the channel width, respectively. Here, we compare our results with the simulations by Feng \etal \citep{FENG1994}, who considered a moderately high Reynolds number ($Re=40$) and a particle with a diameter comparable to the channel width ($D_p/H=0.25$). We consider a square computational domain, where $x$ is the streamwise direction with periodic boundary conditions and $y$ the wall-normal direction with no-slip and no-penetration boundary conditions. The two parallel walls move in opposite directions with the same speed $U_{\text{wall}}/2$ such that the resulting Reynolds number $Re=\rho^f U_{\text{wall}} H / \mu^f$ is equal to $40$. The immersed particle is neutrally buoyant, \ie $\rho^p/\rho^f=1$, and is initially positioned at $y/H=0.25$. The fluid is initialised with the linear profile $u \left( y \right)=U_{\text{wall}} \left( y/H-1/2 \right)$, while the particle is at rest. In \figref{a}{fig:feng} we show the trajectories of the particle center for four different grid resolutions $D_p/\Delta$, $8$, $16$, $32$, $48$, as well as the reference result by Feng \etal \citep{FENG1994}. We observe that the particle migrates towards the center of the channel and a good agreement with the literature results is evident. The results converge to the reference one as the resolution becomes high and the correct result is recovered for $D_p/\Delta$ equal to $16$.

\begin{figure}[t]
\includegraphics[width=\textwidth]{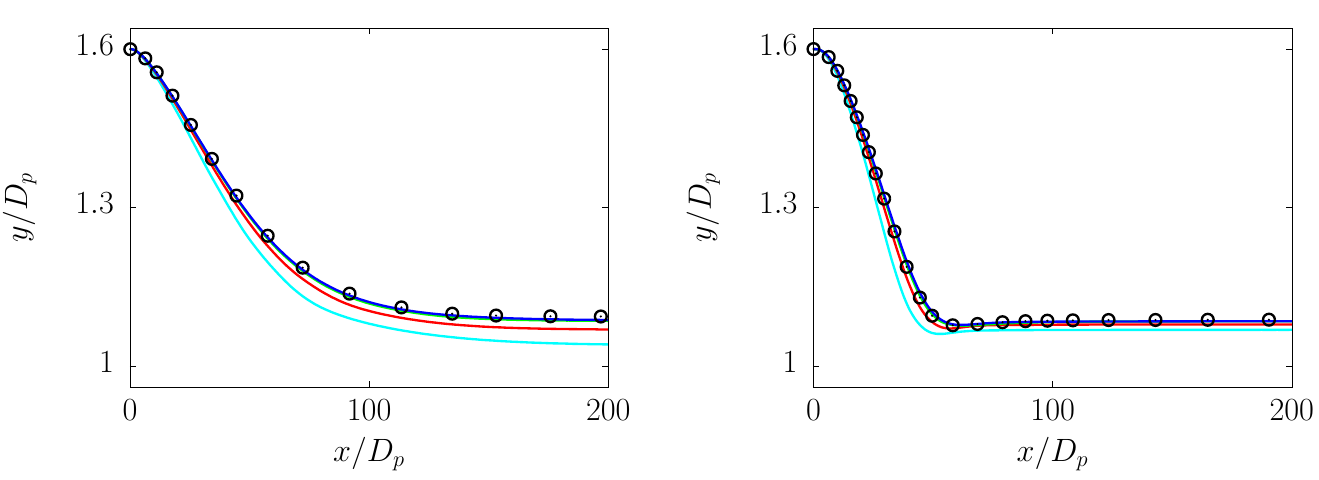}
  \caption{
    The particle trajectory in a pressure-driven flow at different Reynolds numbers: (left) $Re=12.78$, (right) $Re=96.74$. The $x$ axis corresponds to the sum of the distance the particle has moved in the streamwise direction, while the $y$ axis the wall-normal coordinate of the particle center. The solid lines are used to report our results, while the points those by Pan and Glowinski \citep{PAN2002}. Different colors are used to distinguish the different grid resolutions $D_p/\Delta$: (cyan) $8$, (red) $16$, (green) $32$, and (blue) $48$.
  }
  \label{fig:SegreSilberberg}
\end{figure}

\begin{table}[t]
  \centering
  \begin{tabular}{l|l|l|l|l|l|l}
    \hline
    $Re$    & $\mu$                  & $-dp^c/dx$              & $y_{\text{term}}$ & $y_{\text{term}}^*$ & $\omega_{\text{term}}$ & $\omega_{\text{term}}^*$ \\ \hline
    $12.78$ & $3.250 \times 10^{-3}$ & $1.763 \times 10^{-3}$ & $0.2715$          & $0.2732$            & $-0.0536$              & $-0.0535$                \\
    $96.74$ & $4.283 \times 10^{-4}$ & $2.337 \times 10^{-4}$ & $0.2711$          & $0.2722$            & $-0.0505$              & $-0.0505$                \\ \hline
  \end{tabular}
  \caption{
    The detailed configurations of the simulations of the Segr\'{e}-Silberberg effect. In the table we report the fluid viscosity and the imposed pressure gradient driving the flow, together with some results of the simulations. In particular, we report the equilibrium wall-normal position $y_{\text{term}}$ and angular velocity velocity $\omega_{\text{term}}$ of the particles center obtained with the grid resolution $D_p/\Delta=48$. The columns marked with $^*$ are the reference results by Pan and Glowinski \citep{PAN2002}.
  }
  \label{tab:SegreSilberbergconfig}
\end{table}

\begin{figure}[t]
\includegraphics[width=\textwidth]{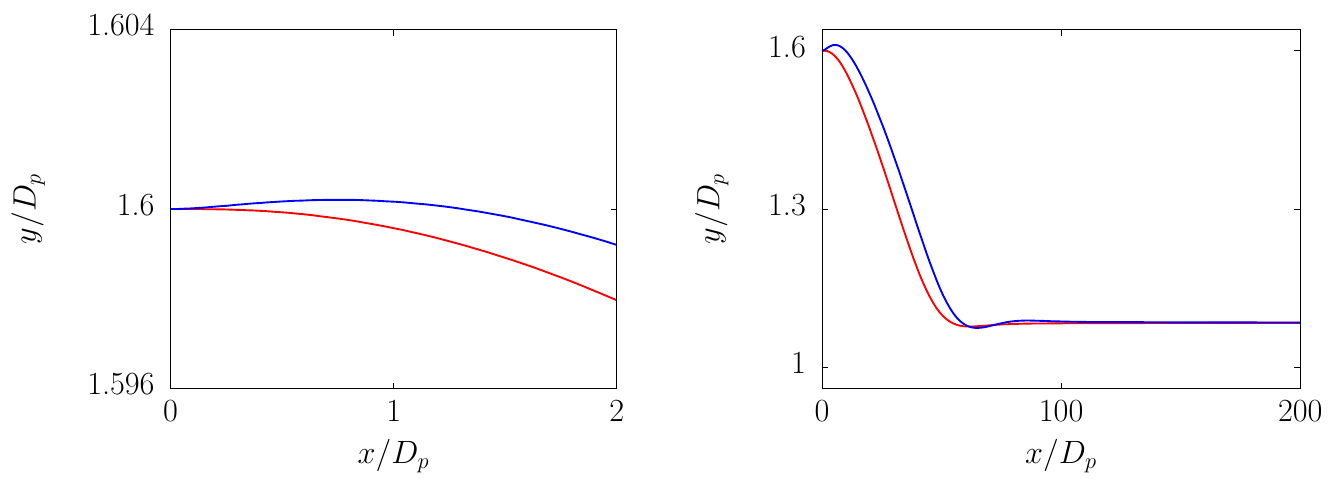}
  \caption{
    The effect of inertia on the particle trajectory at different Reynolds numbers: (left) $Re=12.78$, (right) $Re=96.74$. The axes are same as in \figrefS{fig:SegreSilberberg} and the grid resolution $D_p/\Delta$ is fixed to $48$. The red and blue lines distinguish the cases with and without the inertia term, respectively.
  }
  \label{fig:SegreSilberberginertia}
\end{figure}

\subsubsection*{Lateral migration of a circular cylinder in a pressure driven flow - the Segr\'{e}-Silberberg effect}
In an inertial pressure-driven channel flow, a suspended particle is driven to a certain stable distance from the walls, due to the so-called tubular pinch effect or Segr\'{e}-Silberberg effect named after Segr\'{e} and Silberberg who firstly observed this phenomena experimentally \citep{SEGRE1962}. Several investigations have been conducted later on \citep{HO1974, MATAS2004} and indeed this phenomena is widely used as a benchmark to check the validity of particle-laden simulations \citep{PAN2002}. We consider a rigid circular particle suspended in a two-dimensional pressure-driven channel flow. The numerical domain is a square of size $L$ equal to $4 D_p$, with $D_p$ being the particle diameter, and $x$ and $y$ are the stream-wise and wall-normal directions, respectively. Periodic boundary conditions are applied in the $x$ direction, while the no-slip and no-penetration conditions are enforced in the $y$ direction. We consider a neutrally-buoyant particle with a density $\rho^p$ equal to the liquid one $\rho^f$. The flow is initially at rest, and the particle center locates at $y=0.4L$. A constant and uniform pressure gradient $dp^c/dx$ is imposed from $t=0$, which drives the flow. Two different flow conditions are simulated, leading to two different bulk Reynolds numbers $Re=12.78$ and $96.74$, where $Re$ is defined based on the terminal bulk stream-wise velocity $\bar{u}$ and the channel width $L$, \ie $Re=\rho^f \bar{u} L / \mu$. The detailed configurations and some results are reported in \tabref{tab:SegreSilberbergconfig}. \figrefSC{fig:SegreSilberberg} shows the trajectories of the particle center for four different grid resolutions $D_p/\Delta$, $8$, $16$, $32$, $48$, as well as the result by Pan and Glowinski \citep{PAN2002}, who simulated the same cases using a fictitious domain method. We observe that the particle migrates to a steady equilibrium position, and our results are in good agreements with the reference results. Also, from the figure we can appreciate that the motion of the particle converges to the reference result as the grid resolution increases. As discussed in the previous chapter, the appropriate evaluation of the fictitious fluid inside the particle is important. In \figrefS{fig:SegreSilberberginertia}, we show the results of the same test cases when the fluid inertia within the particle is neglected. Although the final equilibrium position is correct, the dynamics of the particle is wrong: initially the particle moves in the opposite direction, thus leading to a different subsequent dynamics. Reasonably, this error is larger when the Reynolds number is higher and inertial term becomes dominant.

\begin{figure}[t]
  \centering
\includegraphics[width=\textwidth]{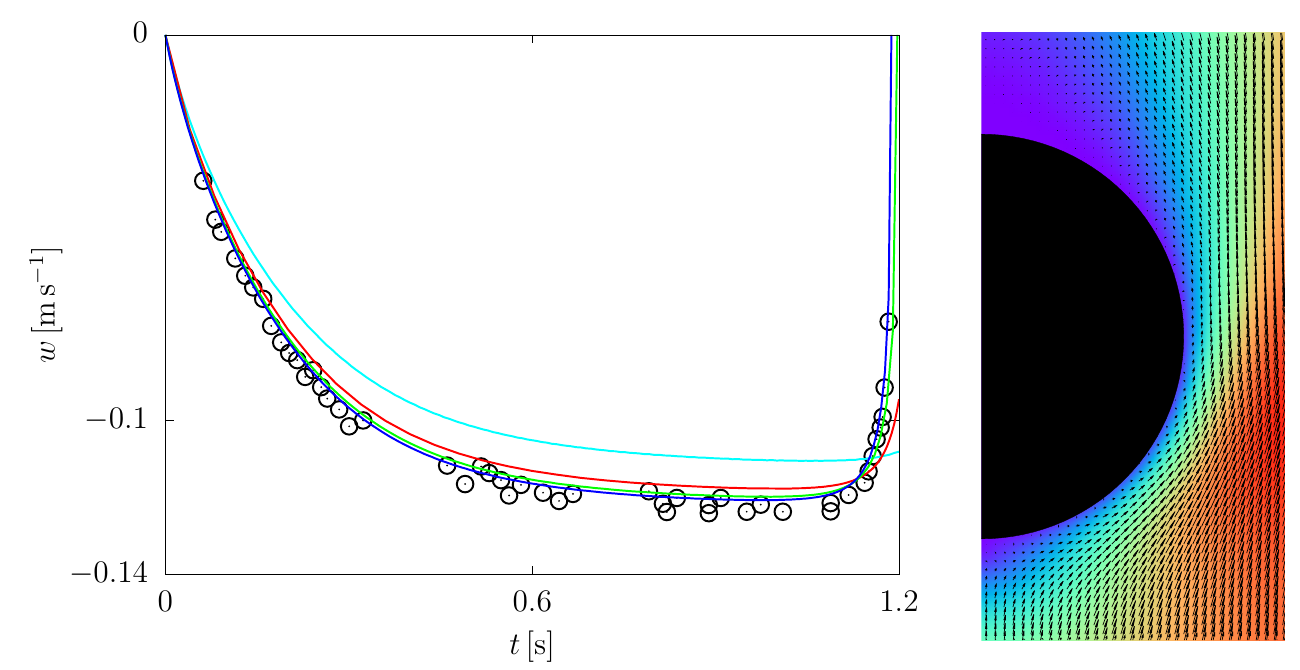}
  \caption{
    (left) The sedimentation velocity of a spherical particle in a closed container. The $x$ axis represents the time after the particle starts moving and the $y$ axis the velocity $w$ in the gravitational direction. The solid lines are used to represent our results and the points the experimental measurements by ten Cate \etal \citep{TENCATE2002}. Different colors are used to distinguish the different grid resolutions $D_p/\Delta$: (cyan) $8$, (red) $16$, (green) $32$, (blue) $48$. (right) Velocity field in the proximity of the falling sphere  for the case with spatial resolution $D_p / \Delta = 48$.
  }
  \label{fig:sedimentation}
\end{figure}

\subsubsection*{Sedimentation of a spherical particle in three dimensions}
Next, we validate our code by checking the sedimentation of a spherical particle. We consider a rectangular domain filled with a liquid with density and viscosity equal to \SI{960}{\kilogram \per \cubic \meter} and \SI{0.058}{\pascal \second}. The domain size is \SI{0.1}{\meter}, \SI{0.1}{\meter} and \SI{0.16}{\meter} in the $x$, $y$ and $z$ directions, where $x$ and $y$ are the two horizontal directions, while $z$ is the vertical direction parallel to the gravitational acceleration $g=$ \SI{9.81}{\meter \per \square \second}. A particle of diameter \SI{0.015}{\meter} and density \SI{1120}{\kilogram \per \cubic \meter} is initially positioned at $\left( \SI{0.05}{\meter}, \SI{0.05}{\meter}, \SI{0.1275}{\meter} \right)$. In \figrefS{fig:sedimentation}, we show the sedimenting velocity $w$ for four different grid resolutions $D_p/\Delta$ equal to $8$, $16$, $32$, and $48$, as well as the experimental results by ten Cate \etal \citep{TENCATE2002}. Our numerical results are in very good agreements with the experiment, thus further showing the validity of our scheme. Note that, when the grid resolution $D_p/\Delta$ is larger than $32$, we are able to capture the sudden slow-down motion of the particle near the bottom wall, which is due to the lubrication effect. Also, the right picture in the figure shows the velocity field in the proximity of the falling sphere, which is smooth without any particular oscillations generated by the particle motion.

\section{The collision and implicit lubrification models}
In this section we move forward our discussion and we consider cases with multiple suspended objects, \ie suspensions, which interact with each other. We will describe two main models that need to be included in the algorithm to properly capture the behaviour of a suspension: the collision and the lubrication models. The former is necessary to model the behaviour of two rigid objects when they are in contact, while the latter to properly capture the force arising as they approach each other. Finally, we will show the validity of our method in a laminar shear flow and in a turbulent pressure driven duct flow laden with rigid spherical particles.

\subsection{Soft-sphere collision model}
\begin{figure}[t]
  \centering
  \includegraphics[width=0.45\textwidth]{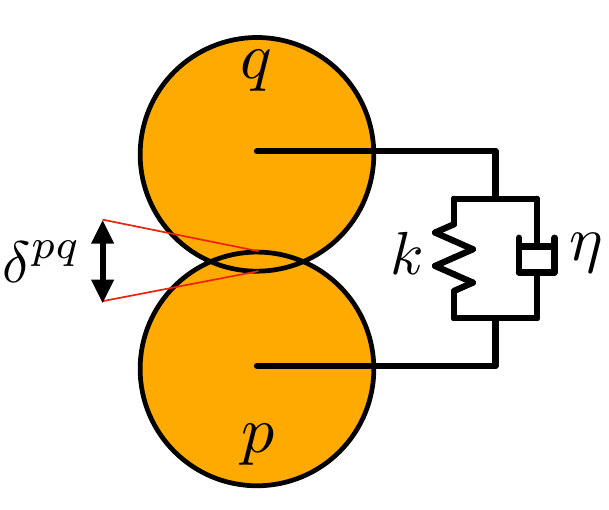}
  \caption{
    The conceptual sketch of the soft-sphere collision model by Tsuji \etal \citep{TSUJI1993}. The gold circles represent the two colliding particles $p$ and $q$. The collision is modeled as a spring and dashpot dynamical system and only the normal collision is considered.
  }
  \label{fig:collision_sketch}
\end{figure}

In order to prevent the interpenetration between particles (or a particle and the wall), we need to introduce a model which adds a repulsive effects. In this section we describe the so-called soft sphere collision model first proposed by Tsuji \etal \citep{TSUJI1993} which we employ in our simulations. In the model, particles are allowed to slightly penetrate and the collision state is spread over multiple time-steps, $T^{\text{col}}$. The collision is modeled as a spring and dashpot dynamical system with the collision force assumed proportional to the penetration depth $\delta^{pq}$ and the relative normal velocity $u_i^{pq}$, where the superscripts $^{pq}$ are used to indicate that we are considering the collision between particle $p$ and $q$ (see \figrefS{fig:collision_sketch}). If we call $n_i^{pq}$ the normal vector from $p$ to $q$, the total force $F_i^{\text{col},pq}$ can be computed as
\begin{equation}
  F_i^{\text{col},pq} = -k \delta^{pq} n_i^{pq} - \eta u_i^{pq},
\end{equation}
where $k$ and $\eta$ are the spring and dashpot coefficients, respectively. These can be found by solving a linear harmonic oscillator problem
\begin{subequations}
  \begin{alignat}{2}
    m^p \frac{d^2x^p}{dt^2} &=  &\eta \left( \frac{dx^q}{dt} - \frac{dx^p}{dt} \right) + k \left( x^q - x^p \right), \\
    m^q \frac{d^2x^q}{dt^2} &= -&\eta \left( \frac{dx^q}{dt} - \frac{dx^p}{dt} \right) - k \left( x^q - x^p \right),
  \end{alignat}
\end{subequations}
with the conditions
\begin{subequations}
  \begin{align}
    &x^p \left( t=0 \right) - x^q \left( t=0 \right) = x^p \left( t=T^{\text{col}} \right) - x^q \left( t=T^{\text{col}} \right), \\
    &\frac{dx^p}{dt} \left( t=0 \right) = v^p, \frac{dx^q}{dt} \left( t=0 \right) = v^q.
  \end{align}
\end{subequations}
The solution of the problem leads to the determination of a unique value of $k$ and $\eta$ which ensures that there is no overlap at the end of the collision, which are
\begin{subequations}
  \begin{align}
    k &= \frac{m^e \left( \pi^2 + \log^2 e_n \right)}{T^{\text{col}}}, \\
    \eta &= -\frac{2 m^e \log e_n}{T^{\text{col}}},
  \end{align}
\end{subequations}
where $e_n$ is the normal restitution coefficient and $m^e$ the harmonic averaged mass.

Apart from the normal collision force discussed above, Tsuji \etal \citep{TSUJI1993} also model a tangential collision force. Costa \etal \citep{COSTA2015} further developed the model and included both the normal and tangential collision forces in the IBM by Breugem \citep{BREUGEM2012}. The procedure to take into account the sliding force is computationally and memory-wise expensive since it requires information from multiple time-steps and multiple particles. Here, we will not include the tangential collision force and we will show that correct results can be obtained in the case of a suspension, thus proving that the tangential collision do not play an important role up to moderately dense regimes (for particle volume fractions less than $50\%$). More details on the topic can be found in the appendix.

The normal collision force is introduced into our immersed boundary method as an external force and we use the Crank-Nicolson scheme to stabilize the time integration. Costa \etal \citep{COSTA2015} suggested the use of a sub-iteration procedure to achieve this, and the same idea can be easily coupled with our immersed boundary scheme by substituting $F_i^{\text{ext}}$ in \equref{eq:ibm_integrate} with $\left( F_i^{\text{ext},k+1}+F_i^{\text{ext},n} \right)/2$. Note that, in the presence of the sole normal collision force, \ie by neglecting the tangential collision force, no additional external torque is added.

The interested reader is referred to \ref{app:bounce} and \ref{app:coll} for more details.

\begin{figure}[t]
\includegraphics[width=\textwidth]{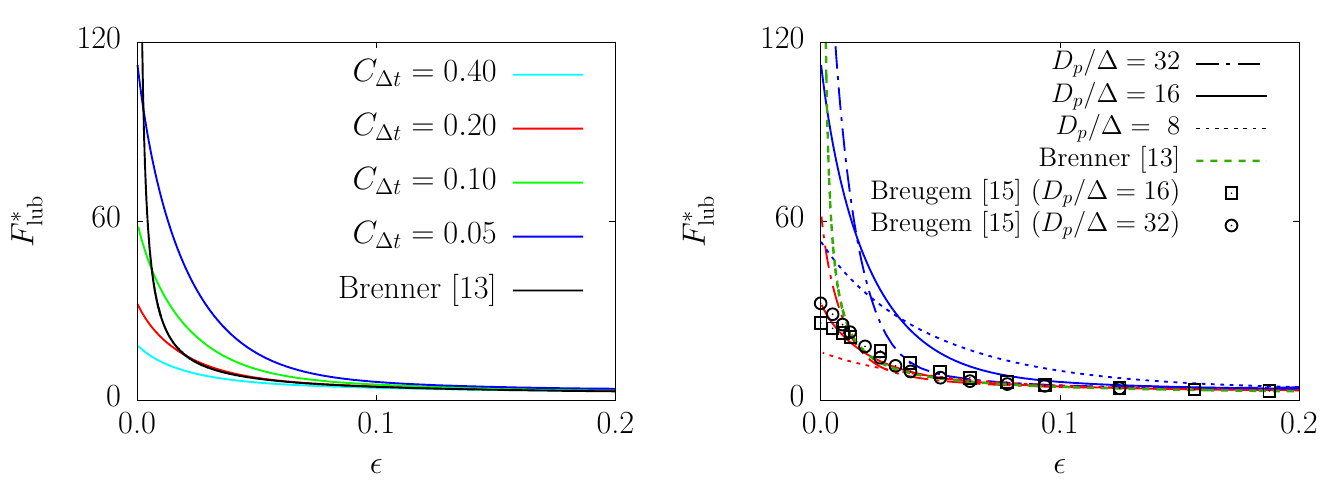}
  \caption{
    The normalised lubrication force $F_{\text{lub}}^*$ as a function of the normalised distance $\epsilon$ of a one-by-one particle interaction. (Left) The effect of the time step. Different colors are used to distinguish different $C_{\Delta t}$: (cyan) $0.4$, (red) $0.2$, (green) $0.1$, and (blue) $0.05$. The black line is the theoretical result by Brenner \citep{BRENNER1961}. In the figure the grid resolution is fixed to $D_p/\Delta=16$. (Right) The effect of the grid resolution. The color scheme is same used in the left figure, while the linestyle represents different grid resolutions $D_p/\Delta$: (dotted) $8$, (solid) $16$, and (dash-dotted) $32$. The black symbols are the numerical results by Breugem \citep{BREUGEM2010}, and the shape represents different grid resolutions $D_p/\Delta$: (square) $16$ and (circle) $8$.
  }
  \label{fig:brenner_result}
\end{figure}

\subsection{The lubrication effect} \label{sec:lub}
We consider two particles frontally approaching each other. Brenner \citep{BRENNER1961} found that two smooth spheres with the same diameter $D_p$ approaching or departing each other with the speed $w$ in an inertialess condition feel a force that counteract their motions, which is called lubrication force $F_{\text{lub}}$. The force can be normalised by the so-called Stokes drag $6\pi\mu wD_p/2$ and the normalised force $F_{\text{lub,theory}}^*$ can be written as
\begin{subequations}
  \begin{align}
    &F_{\text{lub,theory}}^* = \frac{4}{3} \sinh \alpha \sum_{n=1}^{\infty} s_n,
    \intertext{where $\alpha=\cosh^{-1} \left( d/D_p+1 \right)=\cosh^{-1} \left( \epsilon+1 \right)$ and}
    &s_n=\frac{n\left(n+1\right)}{\left(2n-1\right)\left(2n+3\right)} \left[ \frac{2\sinh\left(2n+1\right)\alpha+\left(2n+1\right)\sinh2\alpha}{4\sinh^2 \left(n+1/2\right)\alpha-\left(2n+1\right)^2\sinh^2\alpha}-1 \right].
  \end{align}
\end{subequations}
Note that, here $d$ is the distance from surface to surface of the particles and not the distance of the particles centers, and that we have defined the normalised distance as $\epsilon=d/D_p$. The force monotonically grows as the distance becomes smaller and finally diverges at $d=0$. We performed numerical simulations of the same problem without any lubrication correction, \ie using the natural lubrication arising from the method, and compare our results with this theoretical result. The computational domain is a cube of length $8D_p$, with solid walls in the $x$ and $y$ directions and periodic boundary conditions imposed in the $z$ direction. Note that, we have verified that the results discussed below are independent of the domain size and boundary conditions used. Two particles are initially placed at rest in the middle of the $x$ and $y$ planes and at a distance of $10/8D_p$ in the $z$ direction, \ie at $\left(4D_p, 4D_p, 4D_p+5/8D_p \right)$ and $\left(4D_p, 4D_p, 4D_p-5/8D_p \right)$. A constant acceleration force $\mp f_z^c$ is imposed on the particles, acting in the $z$ direction with the particles approaching each other. We fix the grid resolution to $D_p/\Delta=16$ and solve the system of equations with a constant time-step $\Delta t$. Its value is found in order to satisfy the numerical stability of the algorithm, which in this low-Reynolds number flow is determined by the viscous constraint; in particular, we can define the coefficient $C_{\Delta t}$ as
\begin{equation}
  C_{\Delta t} = \frac{6 \mu^f \Delta t}{\rho^f \Delta^2},
\end{equation}
and determine $\Delta t$ by fixing a sufficiently small value of $C_{\Delta t}$. In \figrefS{fig:brenner_result}, we show the normalised lubrication force $F_{\text{lub}}^*$ as a function of the normalised distance $\epsilon$ for different values of $C_{\Delta t}$, \ie $C_{\Delta t}=0.4$, $0.2$, $0.1$, and $0.05$ and for different grid resolution $D_p/\Delta$, \ie $D_p/\Delta=8$, $16$, and $32$. The analytical solution by Brenner \citep{BRENNER1961} is also reported, together with the numerical results by Breugem \citep{BREUGEM2010}. Note that, the latter obtained the curves by fixing the particle positions at a certain distance and measuring the resulting force, and repeating the procedure for various distances. On the other hand, we allow the particles to move and obtain the full curve with a single run; note also that, we have tried both procedure and indeed found negligible differences in the results. The results in \figref{a}{fig:brenner_result} show that when the distance between the two particles is large, the numerical results provide the correct solution regardless of the $C_{\Delta t}$ used, while the numerical and theoretical results deviate when the distance is small. Also, we observe that when $C_{\Delta t}$ is large, the numerical results mostly underestimate the force but when $C_{\Delta t}$ is small, the force starts growing earlier than in the theory, then it catches up to the correct value and finally the force is again underestimated at contact. Note that, the force is actually converging to the right results as $C_{\Delta t} \rightarrow 0$ at contact ($\epsilon = 0$) but in order to do so at intermediate $\epsilon$ the error has a non-trivial behaviour. The numerical results also depend on the grid resolution as shown in \figref{b}{fig:brenner_result} where we report the results obtained with three grid resolution for two different values of $C_{\Delta t}$. As the resolution of the particle is improved, (\ie $D_p/\Delta$ grows) more of the short-range viscous lubrication forces will be captured directly by the simulation.  However, similarly to what found for the time-step size, also in the case of the grid-size $D_p/\Delta$ we found that the numerical results both underestimate and overestimate the value of the theoretical lubrication force depending on the considered range of $\epsilon$ and that while the convergence is monotonic at $\epsilon=0$, it has a non-trivial behaviour for intermediate $\epsilon$ values.

The complex behaviours discussed above originate from the nature of forcing used to describe the immersed object and can be explained as follows.  The direct forcing introduces an impulse at each time step that captures the hydrodynamic force on the particle, with the time step acting as a penalty parameter to enforce the no-slip condition; this impulse will create a viscous Stokes layer at the boundary of the immersed object,  and refining the time step can thus improve the no-slip condition. However, as the time-step is reduced,  the scale of the layer reduces too,  eventually becoming smaller than the spatial resolution of the simulation. This behaviour was first found and discussed by Luo \etal \citep{luo_maxey_karniadakis_2009i}. Note also that, the parameter $C_{\Delta t}$ we use is roughly the square of the ratio of the Stokes layer scale and the grid resolution \citep{luo_maxey_karniadakis_2009i}.  Finally, the reason why for small time steps the lubrication force is first overestimated for larger gaps and then underestimated for very small gaps is due to the smoothed nature of the interface: for large gaps, the present of the object is felt in advance and thus the lubrication force is overestimated, while for small gaps the force is underestimated due to the finite size of the grid.

A similar trend was found also by Breugem \citep{BREUGEM2010} and Costa \etal \citep{COSTA2015} but not fully discussed: indeed, these authors limited their analysis to large values of $C_{\Delta t}$ which are preferable from a computational point of view, thus resulting in an underestimation of the force in most cases. To correct the underestimated lubrication force, Breugem \citep{BREUGEM2010} suggested to add a corrective force $\Delta F_{\text{lub}}^*$ that increases the under resolved value found naturally and restores the theoretical one, with the correction term $\Delta F_{\text{lub}}^*$ reducing as the resolution of the particle increases, since more of the short-range viscous lubrication force is naturally captured by the simulation. This method was later further developed and verified by Costa \etal \citep{COSTA2015}. Although their correction has been successfully used in the past by several authors, it may lead to an over-estimation of the force in cases where the time-step is reduced due to additional time-constraint. Also, we find that the correction as a tendency to over-estimate the theoretical lubrication force. In their method, the corrective force $\Delta F_{\text{lub}}^*$ is computed as
\begin{equation}
  \Delta F_{\text{lub}}^* \left( \epsilon \right) = F_{\text{lub,theory}}^* \left( \epsilon \right) - F_{\text{lub,theory}}^* \left( \epsilon_0 \right),
\end{equation}
where $\epsilon_0$ is the threshold below which the correction is applied. Since $\Delta F_{\text{lub}}^*$ is added to the force obtained numerically without the correction $F_{\text{lub,no-cor}}^*\left( \epsilon \right)$, the resulting total lubrication force $F_{\text{lub,cor}}^*\left( \epsilon \right)$ included in the numerical scheme is
\begin{equation}
  \begin{aligned}
    F_{\text{lub,cor}}^* \left( \epsilon \right) &= F_{\text{lub,no-cor}}^* \left( \epsilon \right) + \Delta F_{\text{lub}}^* \left( \epsilon \right)= \\
                                                 &= F_{\text{lub,no-cor}}^* \left( \epsilon \right) + F_{\text{lub,theory}}^* \left( \epsilon \right) - F_{\text{lub,theory}}^* \left( \epsilon_0 \right).
  \end{aligned}
\end{equation}
The deviation from the theory $\delta F_{\text{lub}}^* \left( \epsilon \right)$ is thus
\begin{equation}
  \delta F_{\text{lub}}^* \left( \epsilon \right) = F_{\text{lub,no-cor}}^* \left( \epsilon \right) - F_{\text{lub,theory}}^* \left( \epsilon_0 \right),
\end{equation}
which is always positive for $\epsilon$ smaller than $\epsilon_0$, \ie the applied lubrication force is always over-estimated. The deviation is actually not very large when the resolution is appropriate, \eg $\delta F_{\text{lub}}^* \left( 0.001 \right)$ is around $10$ for the grid resolution $D_p/\Delta=32$ which results in an over-estimation of only around $5\%$, but the error becomes larger as the resolution is increased without changing $\epsilon_0$.  Furthermore, in the case of suspensions multiple objects may interact simultaneously, the addition of the external correction force may affect the stability of the method, thus changing the value of $\Delta t$ appreciably and the consequent lubrication force as well.

In the present work, we focus our attention on the non-trivial behaviour of the convergence of the force with $\Delta t$ and $\Delta$ and suggest a methodology to exploit it in order to properly model the lubrication force acting in a suspension. Note that,  our aim in the next section is to provide an easy and simple approach to account for lubrication forces in particle suspensions at finite (both low and high) inertia.  The theoretical estimates of lubrication forces, such as the one discussed above, are all based on viscous Stokes flow; however, if the particle inertia is finite and large,  the short-range hydrodynamic interactions are more rapid and can create finite $Re$ responses. In this case, the simulation results would be more reliable naturally, up to the point the gap is so small that viscous effects dominate again.  Furthermore,  when moderately dense suspensions are considered, the accurate prediction of the motion and forces generated by a single particle becomes less critical since it is the bulk statistical effect that matters.

\begin{figure}[t]
  \centering
  \includegraphics[width=0.6\textwidth]{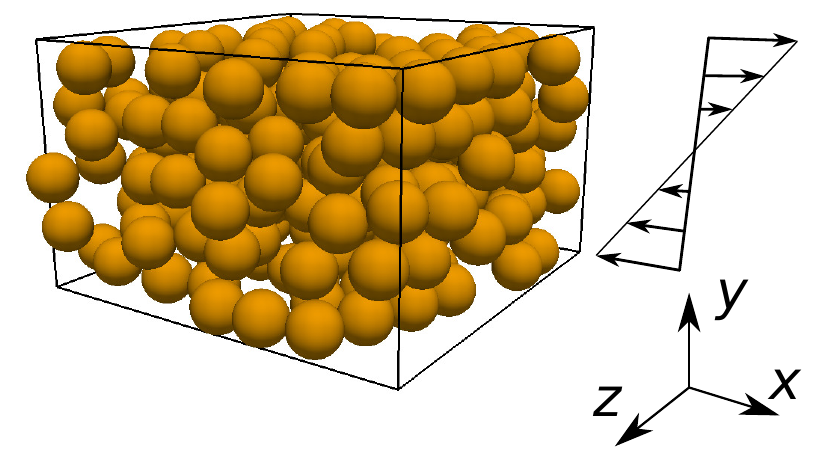}
  \caption{
    The sketch of the system discussed in \secref{sec:eilers} and its reference coordinate system. The gold spheres represent the immersed rigid particles, with a volume fraction $\Phi^p$ equal to $46\%$.
  }
  \label{fig:eilers_sketch}
\end{figure}

\begin{figure}
\includegraphics[width=\textwidth]{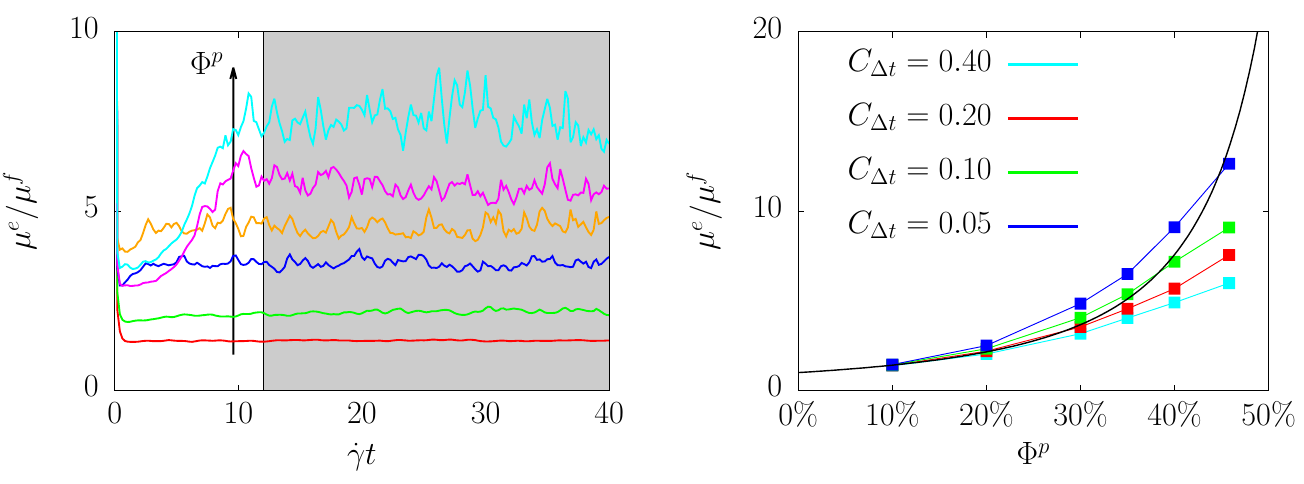}
  \caption{
    (Left) The normalized effective viscosity $\mu^e/\mu^f$ as a function of the normalized time $t^*$. The different colors are used to distinguish different particle volume fractions $\Phi^p$: (red) $10\%$, (green) $20\%$, (blue) $30\%$, (orange) $35\%$, (magenta) $40\%$, and (cyan) $45.8\%$. The results are obtained with $C_{\Delta t} = 0.2$. The gray background indicates the area that we used to compute the statistics. (Right) The normalized effective viscosity $\mu^e/\mu^f$ as a function of the particle volume fraction $\Phi^p$. The different colors are used to distinguish different values of $C_{\Delta t}$: (cyan) $0.4$, (red) $0.2$, (green) $0.1$, and (blue) $0.05$. The black solid line is the Eilers fit of \equref{eq:eilers_fit}, with the fitting parameters $B=1.6$ and $\Phi^{p}_{\text{max}}=0.63$.
  }
  \label{fig:effvis}
\end{figure}

\subsection{The implicit lubrication model based on the rheology of a suspension} \label{sec:eilers}
Here, we consider the rheology of a particle suspension at low Reynolds numbers, \ie when the inertial effect is negligible, in order to explain our implicit lubrication method. We consider a simple Couette flow laden with rigid spherical particles, as sketched in \figrefSC{fig:eilers_sketch}. The $x$, $y$, and $z$ directions correspond to the stream-wise, wall-normal and span-wise directions, respectively, and the domain size is equal to $l_x=8D_p$, $l_y=5D_p$, $l_z=8D_p$, where $D_p$ is the particle diameter. Periodic boundary conditions are applied in the $x$ and $z$ directions, while the no-slip and no-penetration boundary conditions in the $y$ direction. The two walls move with the same speed $U^{\text{wall}}$ towards opposite $x$ directions and the resulting Reynolds number $Re=\rho^f \left( D_p/2 \right) \dot{\gamma}^2 / \mu^f$ is equal to $0.1$, where $\dot{\gamma}$ is the nominal shear rate $2 U^{\text{wall}} / l_y$. The flow is initially at rest and the walls are moved suddenly at $t=0$. The flow is laden with rigid spherical neutrally-buoyant particles; they are initialized randomly in the domain when their volume fraction $\Phi^p$ is relatively small, \ie $\Phi^p \leq 35\%$ while when $\Phi^p$ is larger than $40\%$, they are positioned in a structured arrangement with the addition of some noise to break the symmetry. In the simulations, the collision time $T^{\text{col}}$ is fixed equal to $8 \Delta t$ and the tangential collision neglected. The reader is referred to the appendix for a discussion on their effect. Each computation required around $10^3$ CPU hours conducted using an Intel Xeon Platinum 8280 system.

In such configuration, it is common to study the effective viscosity $\mu^e$ of the suspension, defined as
\begin{equation}
  \frac{\mu^e}{\mu^f} = \frac{\sigma_{xy}^{\text{wall}}}{\mu^f \dot{\gamma}}.
\end{equation}
where $\sigma_{xy}^{\text{wall}}$ is the $xy$ component of the stress tensor at the walls. Einstein \citep{EINSTEIN1911} first derived an analytical expression for $\mu^e$ showing a linear relation with $\Phi^p$ in the case of isolated particles, \ie
\begin{equation}
  \frac{\mu^e}{\mu^f} = 1 + 2.5 \Phi^p,
\end{equation}
and Batchelor and Green \citep{BATCHELOR1972} extended the work and obtained a second order relation to take into account dilute regimes
\begin{equation}
  \frac{\mu^e}{\mu^f} = 1 + 2.5 \Phi^p + 5.2 {\Phi^p}^2.
\end{equation}
These relations are valid in a very limited range of volume fractions and no general theoretical extension exist for dense regimes. Thus, it is common to use empirical formulae to estimate the rheology of the suspension over a wide range of volume fraction, such as the so-called Eilers fit
\begin{equation} \label{eq:eilers_fit}
  \frac{\mu^e}{\mu^f} = \left( 1 + \frac{B \Phi^p}{1-\Phi^p/\Phi^{p}_{ \text{max}}} \right)^2.
\end{equation}
$B$ is a fitting parameter and $\Phi^{p}_{\text{max}}$ the maximum volume fraction of the particles (usually between $0.6$ and $0.65$ for spheres, see \eg the review by Guazzelli and Pouliquen \citep{GUAZZELLI2018}).
In \figref{a}{fig:effvis}, we show typical time-histories of the suspension effective viscosity $\mu^e/\mu^f$ for various volume fractions $\Phi^p$ ($\Phi^p=10$, $20$, $30$, $35$, $40$, and $45.8\%$); for all the volume fractions, $\mu^e/\mu^f$ exhibit an initial transient behaviour and eventually reaches a statistically steady state. The mean values of $\mu^e/\mu^f$ are reported in \figref{b}{fig:effvis} for all the volume fractions considered and for four different $C_{\Delta t}$ equal to $0.4$, $0.2$, $0.1$, and $0.05$. The figure also reports the Eilers fit  (\equref{eq:eilers_fit}) with $B=1.6$ and $\Phi^{p}_{\text{max}}=0.63$. Independently from the value of $C_{\Delta t}$, all our results qualitatively capture the general tendency, with $\mu^e/\mu^f$ monotonically increasing with $\Phi^p$; however, different values of effective viscosity are obtained when varying $C_{\Delta t}$. In particular, the bigger $C_{\Delta t}$ is the smaller the effective viscosity becomes. Although the difference in effective viscosity is small for $\Phi^p=10\%$ and $20\%$, the deviations from the Eilers fit are considerable for more dense regimes. In general however, we found that there is one specific value of $C_{\Delta t}$ which provides the correct rheology of the suspension.

\begin{figure}[t]
\includegraphics[width=\textwidth]{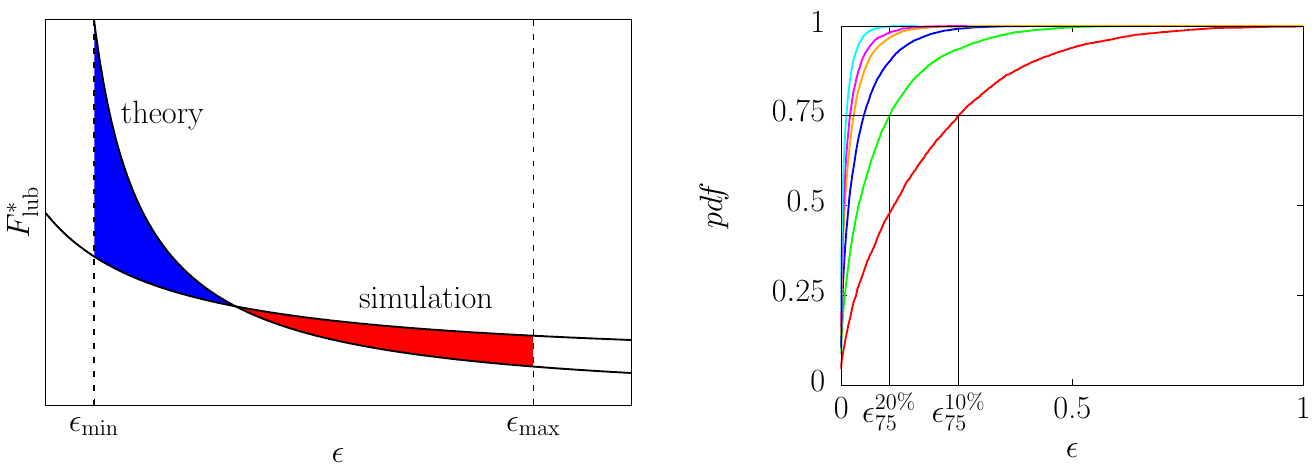}
  \caption{
    (Left) The conceptual sketch of the integration procedure. The two black solid line represent the theoretical curve by Brenner \citep{BRENNER1961} and our numerical results. The blue filled area corresponds to region where the lubrication force is under-estimated, while the red filled area where it is over-estimated. (Right) The cumulative probability density function of the nearest-neighbour distance of all particle pairs. Different colors are used to distinguish different particle volume fraction $\Phi^p$, same as in \figref{a}{fig:effvis}: (red) $10\%$, (green) $20\%$, (blue) $30\%$, (orange) $35\%$, (magenta) $40\%$, and (cyan) $45.8\%$. $\epsilon_{75}^{10\%}$, $\epsilon_{75}^{20\%}$ are two examples of the $\epsilon$ values for which $75\%$ of the cumulative $pdf$ is reached, \ie the right bound of the integration in \equref{eq:integral} and \equref{eq:integral_brenner}.
  }
  \label{fig:pdf}
\end{figure}

Based on these results, our interest is how to determine the appropriate value of $C_{\Delta t}$ which can correctly describe the suspension dynamics without any additional force. Since the collision force is only slightly affected by the choice of the time-step or of the collision time (see the appendix and \cite{COSTA2015}), we can relate the behaviour of $\mu^e/\mu^f$ with $C_{\Delta t}$ to the under- and over-estimation of the lubrication force previously discussed in \secref{sec:lub}. Also, as previously mentioned, the complex behaviour observed for the lubrication force is due to the way in which the no-slip condition is imposed in the immersed boundary method \cite{luo_maxey_karniadakis_2009i}; thus, here we look at the optimal value of $C_\Delta$ that trades off the errors between the increased accuracy coming from an increased time-step and the limit imposed by the spatial grid size resolution. To do that, we combine the information of the particle rheology in \figrefS{fig:effvis} with those of the one-by-one interaction between two particles discussed in \figrefS{fig:brenner_result}. The main difference between the two cases is that in the suspension, multiple particles are interacting at the same time, with each particle pair at a different distance, \ie with a different value of $\epsilon$. The distribution of $\epsilon$ depends on the value of $\Phi^p$, with dilute regimes exhibiting larger distances among particles than in dense regimes. Thus in the simulations, some particle pairs will experience a lubrication force larger and other smaller than the correct one depending on their value of $\epsilon$ once $C_{\Delta t}$ is fixed (the natural lubrication curve captured by the method fixed). What we are aiming for here is to find a value of $C_{\Delta t}$ for which the under- and over-estimation balances out in statistical terms. Mathematically, we can express this concept as the fact that the integral of the lubrication force in the numerical simulations is equal to the integral of the theoretical prediction, \ie the integral of the natural lubrication force captured in the simulations
\begin{equation}
  \mathcal{I}=\int_{\epsilon_{\text{min}}}^{\epsilon_{\text{max}}} F_{\text{lub}}^* d \epsilon, \label{eq:integral}
\end{equation}
should equal the one from theory
\begin{equation}
  \mathcal{I_{\text{theory}}}=\int_{\epsilon_{\text{min}}}^{\epsilon_{\text{max}}} F_{\text{lub,theory}}^* d \epsilon. \label{eq:integral_brenner}
\end{equation}
The best value of $C_{\Delta t}$ is thus the one providing a null error
\begin{equation} \label{eq:integral_error}
  \Delta \mathcal{I} = \mathcal{I}-\mathcal{I}_{\text{theory}}.
\end{equation}
\naokih{
  Note that, \equref{eq:integral} is obtained as an ensemble average of all particle-particle interactions and thus takes into account different approaching velocities, not only different relative distances.}
In \figref{a}{fig:pdf}, we show a conceptual sketch of the process. In the blue filled part, the theoretical curve is above our numerical result, which means that we are under-estimating the lubrication force, while in the red filled part, the behaviour is opposite and we are over-estimating the lubrication force. If these two regions have the same area, then the overall lubrication force is correct on average. In the previous formulae, we have introduced the two extrema of integration $\epsilon_{\text{min}}$ and $\epsilon_{\text{max}}$ which are necessary to take into account the dependency of the particle distribution with the volume fraction as will be discussed next.

\begin{figure}[t]
\includegraphics[width=\textwidth]{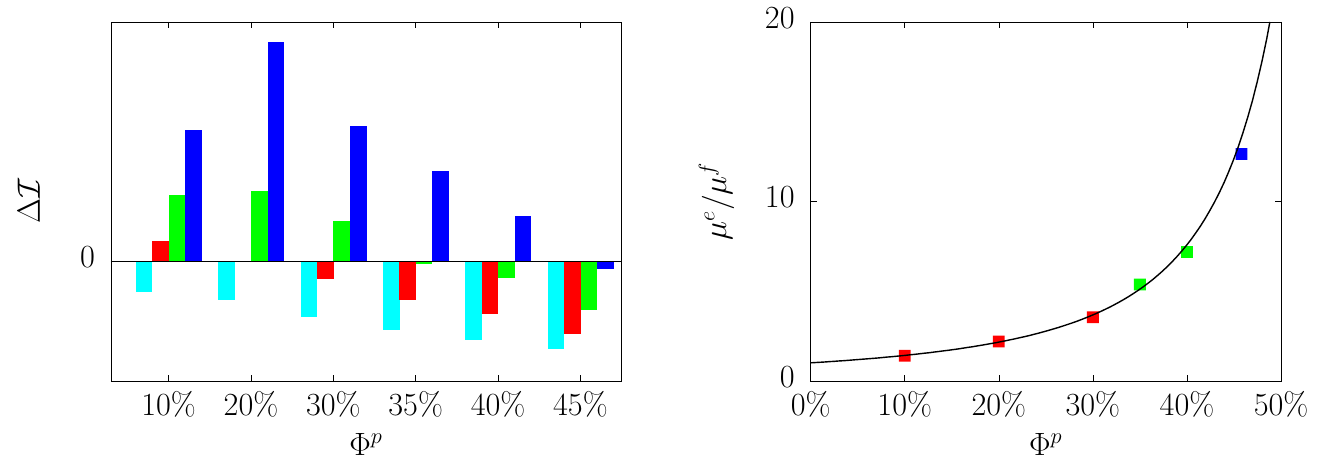}
  \caption{
    (Left) The deviation of the numerical integral from the theoretical one, \ie \equref{eq:integral_error}, as functions of $\Phi^p$.
    Different colors are used to distinguish the different $C_{\Delta t}$, same as \figref{b}{fig:effvis}: (cyan) $0.4$, (red) $0.2$, (green) $0.1$, and (blue) $0.05$.
    (Right) Same as \figref{a}{fig:effvis}, but we now only show the result for the most suitable $C_{\Delta t}$ for which $\left| \Delta \mathcal{I} \right|$ is minimum.
  }
  \label{fig:integral}
\end{figure}

In \figref{b}{fig:pdf}, we report the probability density function ($pdf$) of the nearest-neighbour distance $d_{\text{min}}^p$, defined as
\begin{equation}
  d_{\text{min}}^p = \min_{q \neq p} \left( d^{pq} \right) \;\;\;\;\;\;\; \textrm{where} \;\;\;\;\;\;\; d^{pq} = \| x_i^p-x_i^q \|-D_p.
\end{equation}
The $pdf$ in the figure is shown for different volume fractions $\Phi^p$ in a cumulative format and we only report the results for $C_{\Delta t}=0.2$ for simplicity, since the general conclusion is the same for other values of $C_{\Delta t}$. Also, we neglected all the pairs whose $\epsilon=d^{pq}/D_p$ exceeds $0.2$, because the particles are far enough and the method is naturally able to capture long-range interactions (see \figref{a}{fig:brenner_result}). As expected, the cumulative $pdf$ is more steep as $\Phi^p$ increases, thus indicating that the particles are more close and the right boundary of the integration in \equref{eq:integral}, $\epsilon_{\text{max}}$, should reflect this information. We propose to find $\epsilon_{\text{max}}$ as the $\epsilon$ value for which the $75\%$ of the cumulative $pdf$ is reached, $\epsilon_{75}$, as shown in \figref{b}{fig:pdf} for two specific volume fractions, $10\%$ and $20\%$. In other words this means that $75\%$ of all the particle-particle interactions happen at a distance below $\epsilon_{75}$ and are thus included in the integration. The left boundary of the integration $\epsilon_{\text{min}}$ is introduced in order to neglect the divergence of the theoretical lubrication force reported in \equref{eq:integral_brenner} as $\epsilon$ approaches $0$, which is instead impossible in the simulation. In reality, the lubrication force does not diverge because of the non-smoothness of the particle surface, \ie roughness, and thus Breugem \citep{BREUGEM2010} and Costa \etal \citep{COSTA2015} proposed to limit the growth of the force including a threshold. Apart from introducing a further parameter in the model, we believe that this should depend on the suspension volume fraction because the effect of the rough differs in a dilute or dense regimes. Thus, for the sake of simplicity and to mimic this dependency, we simply fix $\epsilon_{\text{min}}$ equal to $0.1\epsilon_{\text{max}}$. The choice of the two boundaries of the integration has obviously some arbitrariness which is unavoidable in any model, but their choice is not critical and the overall algorithm remains the same when choosing different values. Indeed, different choice of these values will lead to finding slightly different optimal time-step, but the overall procedure and their link to the volume fraction of the suspension remain unchanged.

We report the values of the integral error (\equref{eq:integral_error}) for six different volume fractions $\Phi^p$ and four different $C_{\Delta t}$ in \figref{a}{fig:integral}. For each volume fraction $\Phi^p$, the integral error $\Delta \mathcal{I}$ monotonically increases with $C_{\Delta t}$ in the considered range, and crosses zero only once. As already said, the preferable value for $C_{\Delta t}$ is the one for which this error is zero, or alternatively for which $\left| \Delta \mathcal{I} \right|$ is minimum. Based on this consideration, we show again in \figref{b}{fig:integral} the effective viscosity $\mu^e/\mu^f$ as a function of the volume fraction $\Phi^p$. In particular, we now report one single result for each volume fraction, the one obtained with the $C_{\Delta}$ that minimise the integral error. All the remaining points are indeed very close to the empirical fit, thus confirming the validity of the proposed methodology.

\begin{figure}[t]
  \centering
  \includegraphics[width=0.75\textwidth]{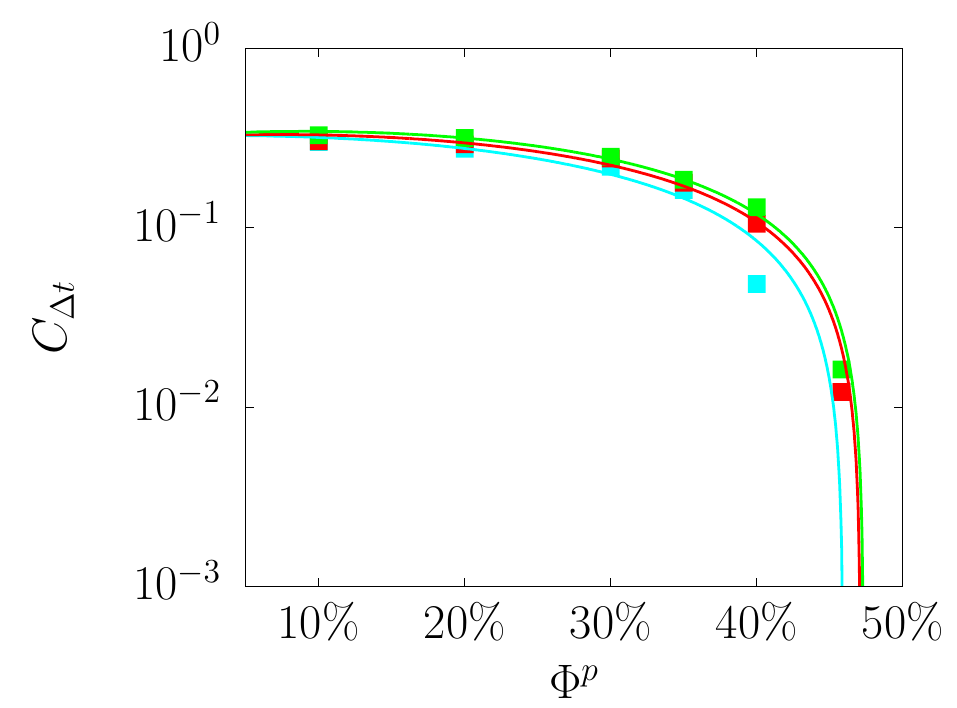}
  \caption{
    The time-step coefficient $C_{\Delta t}$ that minimize the integral error $\left| \Delta \mathcal{I} \right|$ as a function of the volume fraction $\Phi^p$ for two different grid resolution $D_p/\Delta$: (cyan) $8$, (red) $16$, and (green) $32$. The square points represent our data, while the solid lines the interpolant obtained by the least squares method.
  }
  \label{fig:formula}
\end{figure}

We conclude our discussion by providing an empirical formula to find the appropriate time-step coefficient $C_{\Delta t}$ for each particle volume fraction $\Phi^p$ and grid resolution $D_p/\Delta$, \ie $C_{\Delta t} = \mathcal{F} \left( \Phi^p , D_p/\Delta \right)$, and to do so, we find the solution of $\Delta \mathcal{I} =0$ using a simple interpolation of our data. \figrefSC{fig:formula} shows the result of the above procedures for all the volume fractions $\Phi^p$ and for the two grid resolution we have discussed above. As already said, the suitable $C_{\Delta t}$ decreases monotonically with the volume fraction $\Phi^p$: at low volume fractions the variation of $C_{\Delta t}$ is small but then at high volume fractions the appropriate time-step coefficient rapidly decreases.  Also, as expected, $C_{\Delta t}$ increases with $D_p/\Delta$, since more of the short-range viscous lubrication forces is directly captured by the simulation when the resolution is improved.  A fit to our data finally results in the formula
\begin{equation}
  C_{\Delta t} \left( \Phi^p , D_p / \Delta \right) = 0.32 + \mathcal{A} \left( D_p / \Delta \right) \Phi^p + \mathcal{B} \left( D_p / \Delta \right) {\Phi^p}^2,
\end{equation}
where
\begin{subequations}
\begin{align}
  \mathcal{A} \left( D_p / \Delta \right) &= -3.15 \left( D_p / \Delta\right)^{-1} +0.51 \\
  \intertext{and}
  \mathcal{B} \left( D_p / \Delta \right) &= 5.69 \left( D_p / \Delta\right)^{-1} -2.49.
\end{align}
\end{subequations}
\naokih{
  Consistently to our results, the formula predicts that the appropriate value of $C_{\Delta t}$ is almost constant at low volume fractions, while it significantly decreases in packed configurations.
  Also, the value of $C_{\Delta t}$ eventually becomes negative for a certain volume fraction, thus suggesting that this is the maximum volume fraction for which the implicit lubrication method is applicable.
  Note that, the previous formula is a simple fit to our results, and a proper combination of coefficients might be different for other immersed boundary methods.
  However, as we observed in \figrefS{fig:brenner_result}, the under/overestimations of the lubrication force occurs as long as we rely on immersed boundary methods of the direct-forcing type.
  Thus, a similar methodology to evaluate an appropriate $\Delta t$ can be adopted for other immersed boundary methods as well.}



\begin{figure}[t]
  \centering
  \includegraphics[width=0.6\textwidth]{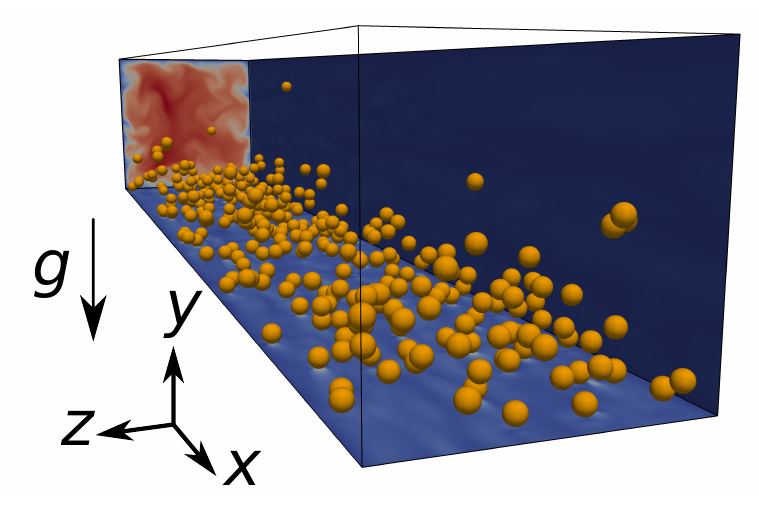}
  \caption{
    The sketch of the system discussed in \secref{sec:turb} and its reference coordinate system. The gold spheres represent the immersed rigid particles, with a volume fraction $\Phi^p$ equal to $1\%$. The three cut planes shows the streamwise velocity contours, with the color scale ranging from $0$ (blue) to $1.4$ (red). Note that, all the planes correspond to one grid cells away from the boundary.
  }
  \label{fig:turb_sketch}
\end{figure}

\subsection{Turbulent duct flow with rigid particles} \label{sec:turb}
In the previous section we described the implicit lubrication model based on an inertialess flow subject to a uniform shear rate, \ie in the absence of shear-induced migration, particle accumulation and particle layering. In order to verify the applicability of the method in the presence of these phenomena, we simulate a pressure driven turbulent square duct flow with particles, and compare our results with the experiments by Zade \etal \citep{ZADE2019}. The chosen configuration presents non-negligible inertia, particle layering at the wall and both shear and gravity-driven migration at the same time and is thus a good test case. A sketch of the computational domain and the reference coordinate system is shown in \figrefS{fig:turb_sketch}; the $x$, $y$, and $z$ directions correspond to the stream-wise and the two wall-normal directions, respectively, and the domain sizes is $l_x=12h$, $l_y=2h$, $l_z=2h$, where $h$ is half the duct size. Periodic boundary conditions are enforced in the $x$ direction, while the no-slip and no-penetration ones in the $y$ and $z$ directions. Particles of size $D_p=2h/14.5$ are suspended in the turbulent flow with a density ratio $\rho^p/\rho^f$ equal to $1.0035$. Gravity acceleration acts in the negative $y$ direction and the resulting Galileo number $Ga= \sqrt{\left( \rho^p - \rho^f \right) g D_p^3/{\mu^f}^2}$ is equal to $40$. Both fluid and particle are initially at rest, and the flow is driven by a time varying pressure gradient that maintains the flow rate constant, resulting in a bulk Reynolds number $\rho^f U_{\text{bulk}} 2h / \mu^f$ equals to $5600$. Two volume fraction $\Phi^p=1\%$ and $3\%$ are considered, and each computation required around $10^4$ CPU hours conducted using an Intel Xeon Platinum 8280 system.

\begin{figure}[t]
\includegraphics[width=\textwidth]{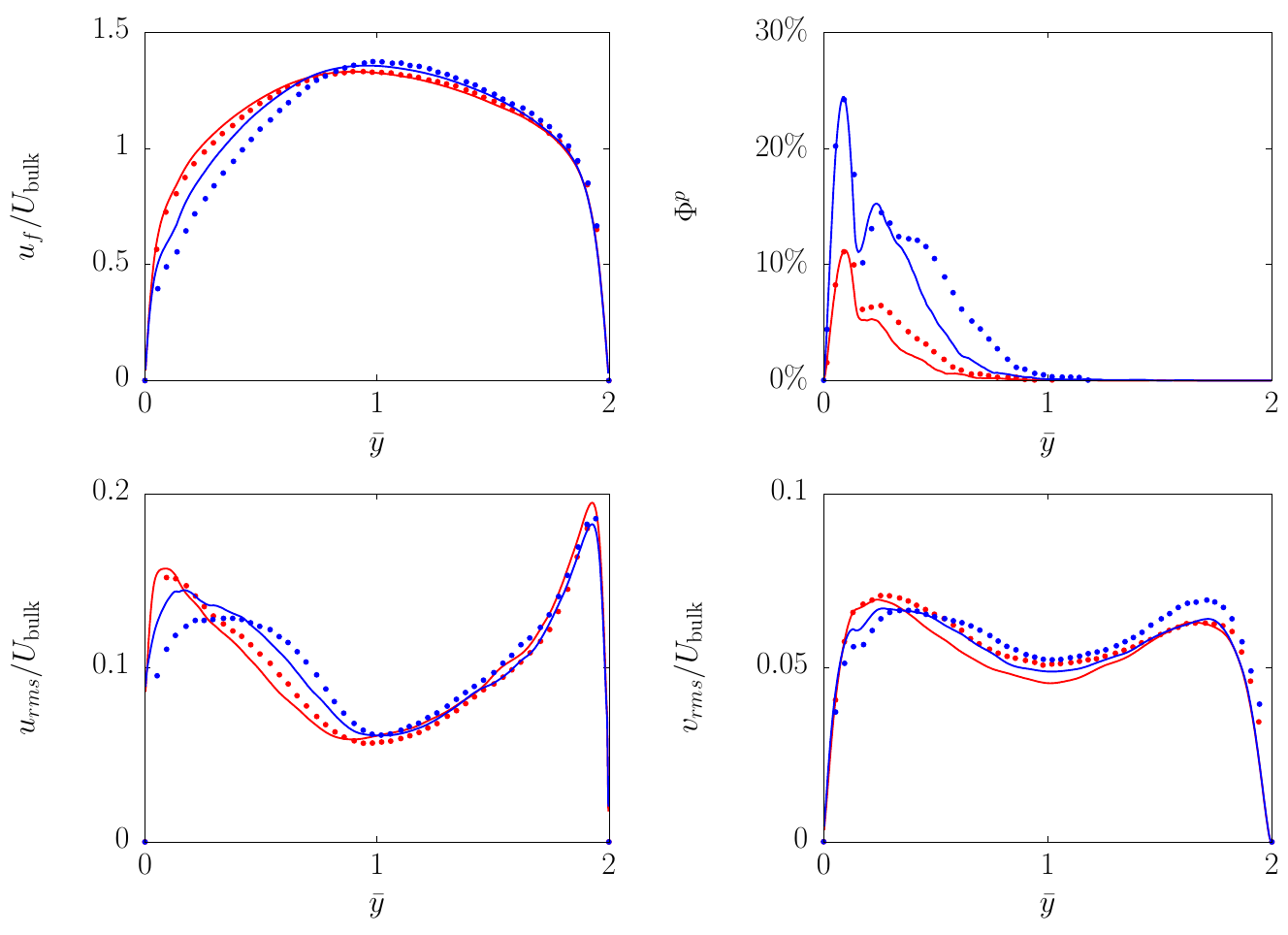}
  \caption{
    Comparison of the turbulent statistic obtained by our simulations (solid lines) and the reference data by Zade \etal \citep{ZADE2019} (points). Different colors are used to distinguish different volume fraction $\Phi^p$: (red) $1\%$ and (blue) $3\%$. (a) the mean stream-wise velocity component, (b) the mean particle concentrations, (c) the $rms$ of the streamwise and (d) wall-normal components of the Reynolds stress tensor, all evaluated in the middle plane of the duct at $z=h$. All velocities are shown as a function of the normalized wall-normal $\bar{y}$ position in bulk units.
  }
  \label{fig:turb_statistics}
\end{figure}

\figrefSC{fig:turb_statistics} shows various turbulent statistics obtained after achieving the statistically steady state; in particular, we report (a) mean stream-wise velocity $u_f=\langle u \rangle_{x,t}$, (b) the mean particle volume concentration $\langle \phi^p \rangle_{x,t}$, the $rms$ of (c) the stream-wise velocity component $u_{rms}=\sqrt{\langle u^2 \rangle_{x,t} - u_f^2}$ and of (d) the wall-normal velocity component $v_{rms}=\sqrt{\langle v^2 \rangle_{x,t} - v_f^2}$ as a function of the normalized wall-normal distance $\bar{y}=y/2h$. Here, $\langle \bullet \rangle_{x,t}$ represents the average operation in the $x$ direction and in time. Good agreement is evident between our results and the reference experiments. Note that, although the total volume fractions are small, due to gravity particles preferentially sediment in the bottom half of the channel and non-dilute local volume fractions (up to $12\%$ for $\Phi^p=1\%$ and $30\%$ for $3\%$) are reached close to the walls.

\section{Conclusions}
We propose an Eulerian-based immersed boundary method (also called Eulerian front capturing method) to simulate rigid objects suspended in a flow. The present methodology is built upon the work by Kajishima \etal \citep{KAJISHIMA2001} and its stability improved for a wide range of particle–fluid mass density ratio, even when unity. This is achieved by accounting for the inertia of the fictitious fluid in the volume occupied by the particles, as first done by Breugem \citep{BREUGEM2012} in a Lagrangian framework. The validity of the method is tested in several benchmarks: we test the particle migration in shear- and pressure-driven flows for neutrally buoyant particles and the gravity-driven sedimentation of a particle immersed in a fluid. Our numerical results are compared with both simulations and experiments from the literature and the method proved to be able to capture the particle dynamics accurately.

Furthermore, we extend the method to the case of suspensions, with multiple interacting suspended particles. We include the soft-sphere normal collision model by Tsuji \etal \citep{TSUJI1993} to prevent the inter-particle penetrations while no additional force is added to correct the subgrid lubrication force. Indeed, the latter is treated implicitly by properly choosing the time-stepping of the numerical simulation: in particular, only one time-step exists that is able to provide the correct macroscopic effect of the lubrication in the suspension and its value is found by minimizing the error between the theoretical lubrication force derived by Brenner \citep{BRENNER1961} for rigid spheres and the natural lubrication captured by the numerical scheme. We show that the time-step providing the correct macroscopic result is a function of the particle volume fraction because of the resulting different particle distributions and we finally provide an empirical fit based on our data that is able to provide the optimum $\Delta t$ as a function of the chosen grid resolution $\Delta$ and volume fraction considered $\Phi^p$.  In particular,  $C_{\Delta t}$ increases with $D_p/\Delta$, since more of the short-range viscous lubrication forces is directly captured by the simulation when the resolution is improved, and reduces with $\Phi^p$. The applicability of this procedure is first tested in a laminar shear flow by studying the rheology of a suspension of rigid spheres, and then in a turbulent pressure-driven flow at high Reynolds number in the presence of non-negligible inertia and non-uniform shear-rate.  The applicability of such technique is based on the fact that we consider suspension flows at finite inertia, where what dominates the results is the bulk statistical effect rather than the accurate prediction of the motion and forces generated by a single particle.

The main advantage of the proposed immersed boundary method is its high efficiency in terms of computational cost which derives by its intrinsic Eulerian nature with the absence of any Lagrangian points: this massively simplifies the numerical scheme, makes its parallelisation procedure straightforward and allows for fast computation and an easy migration towards the rapidly growing GPU computations. Furthermore, only the addition of a normal soft-sphere collision model is needed to properly handle full suspensions, since the correction of the subgrid lubrication force is treated implicitly. The new method has the additional advantage that any shape can be treated easily with only minor modifications, also thanks to the simple and fast digitaliser by Yuki \etal \citep{YUKI2007} which evaluates the local particle volume fraction by assuming a sigmoid-like surface at the interface. This is similar to what usually done in other Eulerian techniques and more consistent when studying multiphase flows with more than two phases.

\appendix
\section{Non-spherical particles}
\begin{figure}[t]
\includegraphics[width=0.9\textwidth]{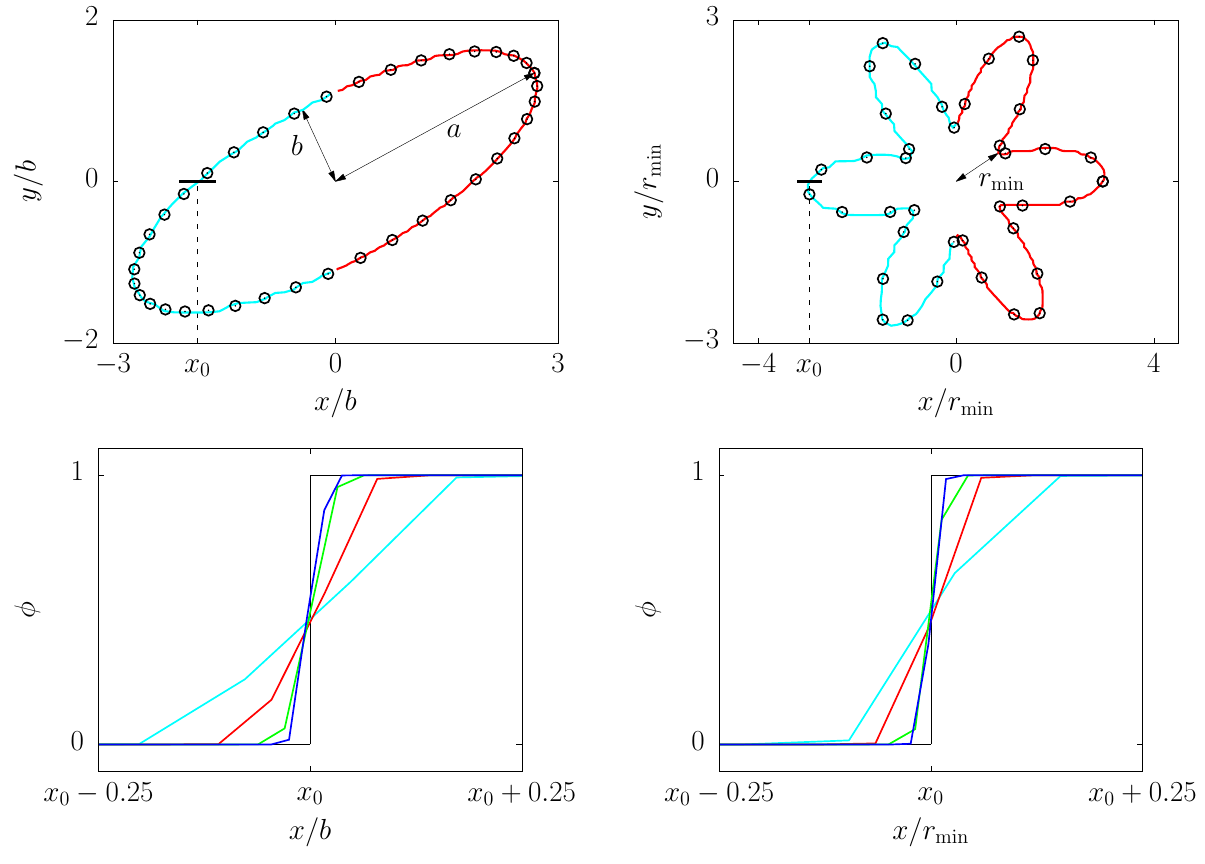}
  \caption{
    The results of the surface digitalisation of non-spherical shapes using the method proposed by Yuki \etal \citep{YUKI2007}. Different line colors are used to distinguish different grid resolutions: (cyan) $8$, (red) $16$, (green) $32$, and (blue) $48$. (Top) The digitalised surfaces of (Left) an ellipse and (Right) a star-like shape compared with the correct analytical results shown with the black circles. (Bottom) The value of the volume fraction $\phi$ as a function of the position on the black lines marked in the upper figures; $x_0$ is the center of the line and $x_0 \pm 0.25$ are each edges. The ellipse resolutions is provided as $b/\Delta$, where $b$ is the semi-minor axis, while the star-like one is based on the shortest distance from the center $r_{\text{min}}/\Delta$.
  }
  \label{fig:ellipse_geo}
\end{figure}

The digitaliser we use proposed by Yuki \etal \citep{YUKI2007} can be easily extended to objects with a non-spherical (or non-circular) shape as long as the normal vector on the surface is properly defined. In this appendix we show some examples of the surface digitalisation of non-spherical objects and its coupling with the proposed IBM .

\subsection*{Surface digitalisation}
We consider two different shapes, an ellipse and a star-like shape, whose normal vectors on the surface can be obtained analytically. The ellipse has the aspect-ratio $a/b=3$, where $a$ and $b$ are the semi-major and minor axes, respectively, and we rotate the shape from the $x$ axis by $\tan^{-1} \left( 1/2 \right)$. The parametric formulation of the star-like shape is $r=1+1/2\cos\left(6 \theta+\pi/12\right)$. In the upper row of \figrefS{fig:ellipse_geo}, we show the surfaces of these shapes, \ie the contour line with the volume fraction $\phi=0.5$. We can see a general good agreements even with lowest resolution considered. In the lower part of the figure, we show the variation of $\alpha$ on the black lines reported in the panels above. We observe that the surfaces become steeper as the resolution is increased and eventually converge to the step function as the grid size goes to zero.

\begin{figure}[t]
  \centering
  \includegraphics[width=\textwidth]{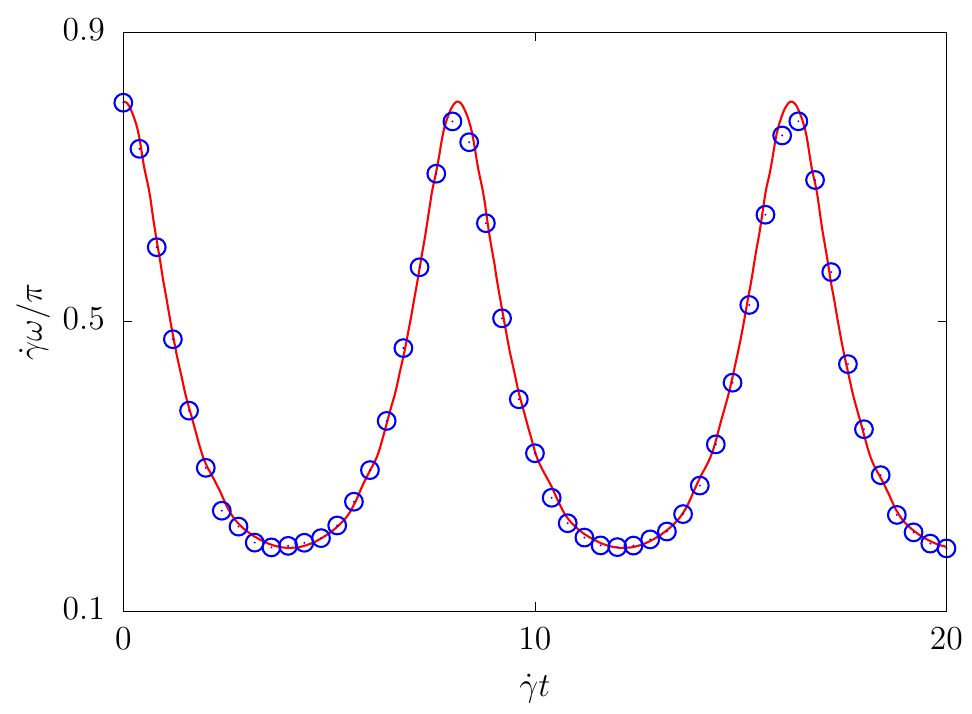}
  \caption{
    The normalised rotational velocity $\dot{\gamma}\omega/\pi$ as a function of the normalised time $\dot{\gamma}t$. The red solid line is used to show our results, while the blue circular points are the reference data by Aidun and Ding \citep{AIDUN1998}.
  }
  \label{fig:ellipse_rotation_vel}
\end{figure}

\subsection*{The rotation of an elliptical object in a shear flow}
We show the validity of the proposed immersed boundary method for non-spherical shape in two dimensions by measuring the rotation of an ellipse rotating in a shear flow. Jeffery and Filon \citep{JEFFERY1922} found an analytical solution for the rotation of an ellipse (or an ellipsoid) in an inertialess shear flows and Aidun and Ding \citep{AIDUN1998} compared their lattice-Boltzmann simulations with the theory in the limit of vanishing inertia, and also reported the results for finite inertia. Here, we compare our results with the latter case by considering an ellipse with aspect ratio $a/b$ fixed equal to $2$ and the confinement ratio with the channel height $a/H$ equal to $0.1$. The computational domain is same we described in \secref{subsubsec:cylinder_in_shear}, but the resulting channel Reynolds number $Re$ is equal to $10$ (the particle Reynolds number $Re_p=1$). The numerical methodology is the same we used for sherical shapes, except in the definition of the mass $m^p$ and moment of inertia $\mathcal{I}^p$ of the ellipse which are equal to $\pi \rho^p ab$ and $m^p \left( a^2 + b^2 \right) / 4$, respectively. In \figrefS{fig:ellipse_rotation_vel}, we report the normalised rotational velocity $\dot{\gamma}\omega/\pi$ of the particle as a function of the normalised time $\dot{\gamma}t$, where we compare our results with those by Aidun and Ding \citep{AIDUN1998}. A very good agreement is found.

Note that, most of the above methodology can be directly adapted also in three dimensions, but we need to consider the full moment of inertia tensor $\mathcal{I}_{ij}$ in \equref{eq:newtoneuler} which is now a function of time due to the rotation, and thus should be positioned inside the time derivative. Furthermore, the application and validation of the implicit lubrication in the case of non-spherical shape will be left for future works.

\begin{figure}
    \centering
    \includegraphics[width=0.5\textwidth]{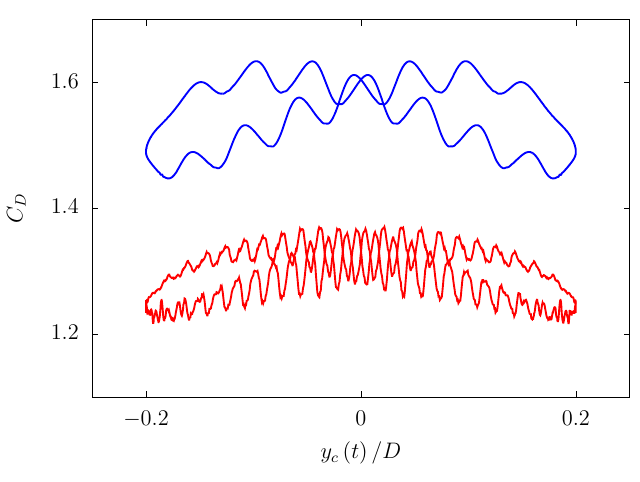}
  \caption{
    Time-periodic variation of the drag coefficient in the case of a translationally oscillating cylinder in uniform cross-flow: (red) standard and (blue) smoothed digitizer.
  }
  \label{fig:forced_cylinder}
\end{figure}

\section{Forced oscillation on a cylinder} \label{app:forced}
Uhlmann \citep{UHLMANN2005} simulated a rigid cylinder oscillating in time in a direction normal to the incoming flow.  The amplitude of the oscillation is set to $0.2D$ and the frequency to $0.8$ times the natural shedding frequency.  The author showed that with his method the drag coefficient time history is slightly oscillating, with the latter reducing by using the regularized delta function of Peskin with wider support. On the other hand, the author also showed that the corresponding result obtained by means of the forcing method of Kajishima \cite{KAJISHIMA2001} produced strong oscillations on the scale of the mesh-width. Moreover, he showed that the smoother the representation of the interface, the higher is the value of the mean drag. For the sake of completeness, we reproduce here the results using the proposed method in a set-up identical to the one used by Uhlmann \citep{UHLMANN2005}; as can be seen in \figrefS{fig:forced_cylinder}, indeed the present method produces strong oscillations in the resulting force coefficient; however, as also noted for the Lagrangian method, the amplitude of the oscillations can be reduced and smoothed out by using different surface digitizer and spreading the interface over a larger number of points. A drawback of this corrective approach, similarly to what found when using a larger support for the delta function in Lagrangian methods,  is that the mean value of the drag coefficient increases; this could be reduced by retracting the interface with an idea similar to the one used by Breugem \cite{BREUGEM2012}, but is out of the scope of the present work. Please note that, these force oscillations are strongly reduced when the particle is let free to move, and amplifies when the motion is imposed.

\begin{figure}
    \centering
    \includegraphics[width=0.5\textwidth]{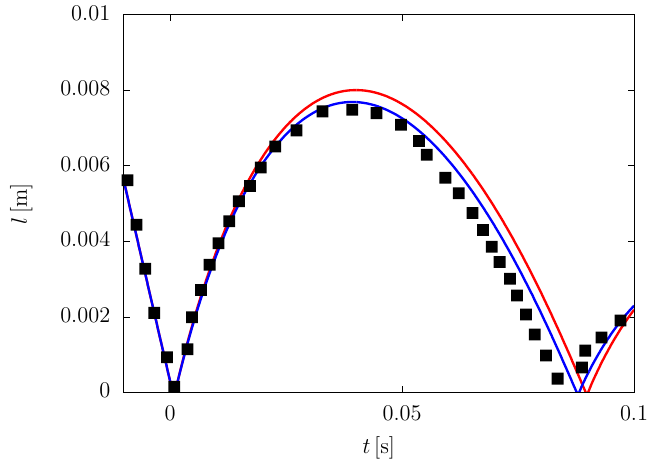}
  \caption{
    Trajectory in the bouncing motion of a sphere colliding onto a planar wall. The points represent the experimental measurements by Gondret. \etal \cite{gondret2002bouncing}, while the red and blue line our numerical results obtained with $C_\Delta=0.5$ and $0.25$.
  }
  \label{fig:bounce}
\end{figure}

\section{Bouncing of a sphere onto a wall} \label{app:bounce}
We simulate the motion of a settling sphere colliding onto a planar wall; the trajectory of the point of the particle closest to the surface is compared to the experimental data of Gondret. \etal \cite{gondret2002bouncing}. This test case was used by Costa \etal \citep{COSTA2015} as a useful for confirming the validity of the lubrication corrections and collision model. As in their work, we consider a numerical domain with dimensions $L_x \times L_y \times L_z = 12D_p \times 30D_p \times 12D_p$; the particle is released from a height of $y =L_y - 0.75D_p$, and the motion is driven by the gravitational acceleration $g = 9.81 m/s^2$.  \figrefSC{fig:bounce} shows the trajectory of the particle bouncing on the wall, with our results compared with the experimental data by Gondret \etal \cite{gondret2002bouncing}. We observe that the collision model implemented in our method successfully produce the qualitative bouncing motion of the sphere; however the results slightly overpredict the particle rebound. Although the error (on a fixed grid) can be reduced by decreasing the value of $C_\Delta=0.5$. we point out that our aim is not to fully and accurately predict the lubrication between two spheres or a single sphere and a wall, but we aim to reproduce the correct macroscopic effect in a suspension in a statistical sense.

\begin{figure}
\includegraphics[width=\textwidth]{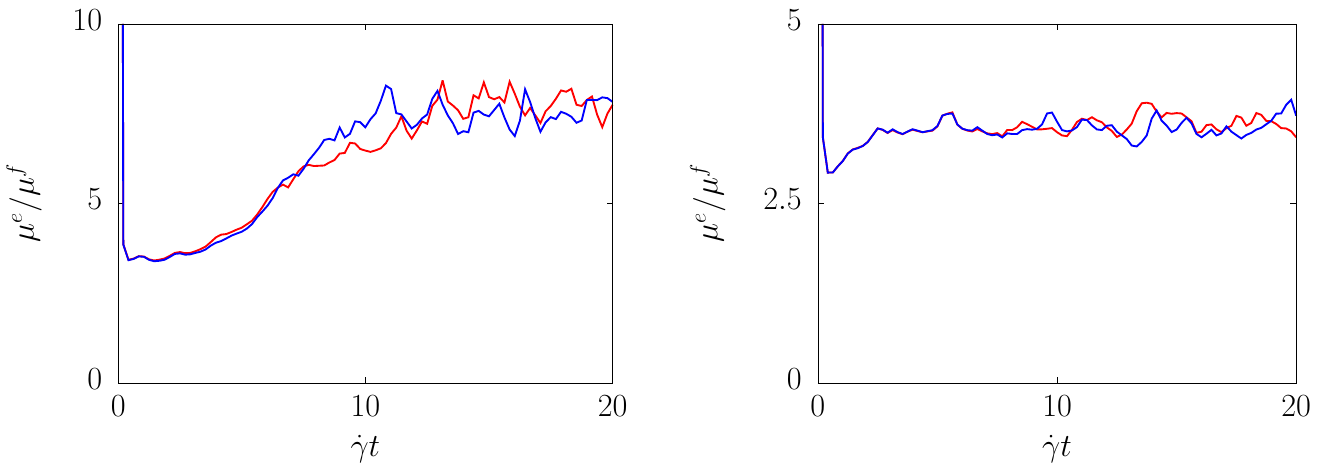}
  \caption{
    The time history of the normalized effective viscosity $\mu^e/\mu^f$ obtained with $C_{\Delta t}$ fixed equal to $0.2$. (Left) The effect of the collision time $T^{\text{col}}$, with the red and blue colors used to distinguish $T^{\text{col}} = 4\Delta t$ and $8\Delta t$. (Right) The effect of the tangential collision, with the red and blue colors used to distinguish the cases with and without the tangential collision.
  }
  \label{fig:effvis_history_appendix}
\end{figure}

\section{The collision time and the tangential collision} \label{app:coll}
\subsection*{The collision time $T^{\text{col}}$}
As discussed in the main text, we use the soft-sphere collision model to model the contact dynamics. In the model, the collision time is stretched over a certain time $T^{\text{col}}$ and here we examine the effect of this parameter on the rheology of a suspension in a dense regime with $\Phi^p$ equal to $45.8\%$. All the set-up is the same used in \secref{sec:eilers} and we only vary the value of $T^{\text{col}}$. In \figref{a}{fig:effvis_history_appendix} we show the time history of the effective viscosity for two different $T^{\text{col}}$ equal to $4\Delta t$ and $8\Delta t$. From the results we observe that, although the time history is obviously different, the statistical steady state value remain unchanged and we can conclude that the effect of $T^{\text{col}}$ on the suspension rheology is limited.

\subsection*{The tangential collision}
Costa \etal \citep{COSTA2015} propose to model the tangential component of the collision as a spring-dashpot system with a friction sliding element. To do this, the algorithm needs to store and accumulate information from the past time-steps with an overload in memory and computational cost. In all the discussion of \secref{sec:eilers} we have always neglected the tangential collision and here we examine its effect on the rheology of a suspension in a semi-dense regime with $\Phi^p$ equal to $30\%$. All the set-up is the same used in \secref{sec:eilers} and we only include the tangential friction model with the tangential restitution coefficient $e_t$ equal to $0.1$ and the friction coefficient $\mu^{\text{frc}}$ equal to $0.15$. In \figref{b}{fig:effvis_history_appendix} we show the time history of the effective viscosity with and without the tangential collision; we observe that altough instantaneous differences, the statistically steady state value of the effective viscosity is unchanged and we conclude that the effect of the tangential component of the collision on the suspension rheology is limited. Note however that, this may not be true for very dense regime; indeed, Mari \etal \citep{MARI2014} reported that the friction plays an important role to reproduce the rapid increase of the effective viscosity when the particle volume fraction is close to the maximum packing, \ie for $\Phi^p$ larger than around $50\%$. From this concentration onward, the modeling of the hydrodynamic effect becomes secondary while the mechanical contacts are the dominant ones.

\section*{Acknowledgments}
MER was supported by the FY2019 JSPS Postdoctoral Fellowship for Research in Japan (Standard), P19054. The authors acknowledge computer time provided by the Supercomputing Division of the Information Technology Center, The University of Tokyo.

\bibliographystyle{ieeetr}

\end{document}